\documentclass[conference,review,10pt]{IEEEtran}
\IEEEoverridecommandlockouts

\usepackage{cite}
\usepackage{amsmath,amssymb,amsfonts}
\usepackage{graphicx}
\usepackage{textcomp}
\usepackage{xcolor}
\usepackage{siunitx}
\usepackage{booktabs}
\def\BibTeX{{\rm B\kern-.05em{\sc i\kern-.025em b}\kern-.08em
    T\kern-.1667em\lower.7ex\hbox{E}\kern-.125emX}}

\usepackage{microtype}
\usepackage{amsmath,amssymb,amsfonts}
\usepackage{graphicx}
\usepackage{textcomp}
\usepackage{color,soul}
\usepackage{listings}
\usepackage[font=footnotesize]{caption}
\usepackage{subcaption}
\usepackage[hyphens,spaces,obeyspaces]{url}

\usepackage[hidelinks]{hyperref}
\usepackage[symbol]{footmisc}

\usepackage{tikz}
\usepackage{multirow}
\usepackage{quantikz}
\definecolor{bg}{RGB}{248,248,248}
\usepackage{adjustbox}

\usepackage{algorithm}
\usepackage{algpseudocode}

\usepackage[inline]{enumitem}

\begin{document}

\title{Dependence-Driven, Scalable Quantum Circuit Mapping with Affine Abstractions
}

\newcommand{\mk}[1]{\textcolor{red}{{(MK):#1}}}
\newcommand{\mkleft}[1]{\textcolor{red}{{$\leftarrow$ (MK) #1}}}
\newcommand{\mb}[1]{\textcolor{blue}{{(MB):#1}}}
\newcommand{\bm}[1]{\textcolor{cyan}{{(BM):#1}}}
\newcommand{\rb}[1]{\textcolor{orange}{{(RB):#1}}}
\newcommand{\ourtool}[0]{{\tt Qlosure}}

\newcommand{\shrink}[1][1]{\vspace{-#1\dimexpr0.25cm\relax}}
\newcommand{\shrinklittle}{\vspace{-0.10cm}}
\newcommand{\TableFontSize}{\scriptsize}
\newcommand{\FormulaSize}{\footnotesize}
\newcommand{\settablefontsize}{\TableFontSize}
\newcommand{\fontsizefortables}{\TableFontSize}
\newcommand{\seteqfontsize}{\FormulaSize}

    \newcommand{\HeuristicCost}{M}

\author{
\\
\IEEEauthorblockN{Marouane Benbetka, Merwan Bekkar}
\IEEEauthorblockA{
\textit{Ecole nationale Supérieure d'Informatique}\\
Algiers, Algeria\\
km\_benbetka@esi.dz, km\_bekkar@esi.dz}
\and
\IEEEauthorblockN{Riyadh Baghdadi}
\IEEEauthorblockA{\textit{Computer Science Program} \\
\textit{New York University Abu Dhabi}\\
Abu Dhabi, UAE\\
baghdadi@nyu.edu}
\and
\IEEEauthorblockN{Martin Kong}
\IEEEauthorblockA{\textit{Dept. of Computer Science } \\
\textit{and Engineering} \\
\textit{The Ohio State University}\\
Columbus, Ohio, USA\\
moreno.244@osu.edu}
}

\maketitle

\begin{abstract}
Qubit Mapping is
a critical task in Quantum Compilation,
as modern Quantum Processing Units (QPUs)  
are constrained to nearest-neighbor interactions
defined by a qubit coupling graph.
This compiler pass
repairs the connectivity
of two-qubit gates whose operands
are not adjacent
by inserting SWAP gates
that move the state of qubits between
directly connected qubits.
Deciding when to introduce SWAPs 
while minimizing their count 
is critical because the error
in quantum programs increases exponentially
with the circuit latency, measured in number
of gates along the critical path of the circuit.
Prior work for this problem relied on
heuristics and exact methods that
partition the circuit into two or more
layers, but failed to exploit valuable dependence
information in any form.

This paper introduces a novel qubit
mapping algorithm based on the weight
of transitive dependences. The introduced 
mapper models quantum circuits with
affine abstractions, thereby providing
the ability to compute transitive dependences.
In turn, the newfound information is used
to partition circuits by dependence distances
and compute, efficiently, distinct weights for each layer. 
We evaluate the efficiency of our mapper
on IBM and Rigetti QPUs, using the large datasets from the 
QUEKO and QASMBench benchmark suites, and against
four baseline tools (QMAP, Sabre, Cirq and TKET),
demonstrating notable improvements in circuit
depth and swap count while delivering
competitive scalability.

\end{abstract}


\thispagestyle{plain}
\pagestyle{plain}

\section{Introduction}
\label{sec:intro}
\newcommand{\qc}[0]{QC}

Quantum Computing (\qc{}) aims to exponentially accelerate
applications in several domains, including complex physical simulations,
hybrid classical/quantum drug design \cite{drug-design.ibm.2018},
cryptography, and others \cite{ibm-quantum-experience.nature.2017}.
In today's era of Noisy Intermediate Scale Quantum (NISQ) devices, quantum programs are written in a device-agnostic fashion.
Quantum compilers then translate and map quantum programs
to target devices.

One of the critical tasks of quantum
compilers is to repair the connectivity
of quantum programs (circuits). 
This problem is known as
{\em qubit mapping} \cite{olsq},
and requires to {\em move} the qubit
state among adjacent, physically
connected qubits, via SWAP operations,
until qubit operands become directly connected.

Finding exact and optimal mapping
for quantum programs is exponentially complex 
\cite{siraichi.cgo.2018}.
Current quantum compilers 
rely either on
graph-driven heuristics 
that perform a few traversals on the input program
\cite{qubit-mapping-hardware-aware.tqe.2020, topo-graph-dependencies-qubit-mapping-heuristic.tqe.2022,siraichi.cgo.2018,tket-placement.dagstuhl.2019,sabre},
or 
on 
computationally expensive and {\em exhaustive} techniques
\cite{olsq,tannu.asplos.2019,jku.date.2018,shi.asplos.2019,qubit-mapping-routing-with-maxsat.micro.2022}
based on Integer Linear Programming and Satisfiability Modulo Theories (SMT) solvers.
The former class of methods are
fast and scalable, but
are limited to making local decisions as the
circuit is traversed, while the ones in the
latter class
are precise, but are limited to 
short
circuits with a few hundred gates and on 
Quantum Processing Units (QPUs) with only a few tens
of qubits.
The commonality between these two methods
is that they model quantum programs at the granularity
of a gate and individual dependences 
among them. Hence, they fail to
exploit regular structures such as the behavior common to several operations and the increasing hierarchical nature of 
quantum devices.
More concretely, prior works fail
to exploit powerful information
of yet-to-map circuit layers in the form
of transitive dependences.

The main scientific contributions of our
work are a methodology and algorithm
to effectively harness the hidden information
in transitive dependences and
tackle the qubit mapping  problem. 
Our mapping
algorithm, \ourtool{},
exploits the information
captured in transitive dependences,
after the program has been lifted
from its QASM representation \cite{openqasm.arxiv.2017}
to {\em affine abstractions} \cite{qrane,kong.tqc.2021}.
Transitive dependences are valuable because
they provide powerful information on the 
effect
of local mapping decisions in later
circuit layers. 
Intuitively, transitive dependences permit
us to evaluate the effects of mapping decisions
on immediate gates and, crucially,
on much farther layers.
More precisely, transitive dependences
computed via polyhedral libraries 
-- e.g., the Integer Set Library (ISL) \cite{isl} --
effectively permit to compute
all possible qubit paths affected by
mapping decisions.
Such information is similar to, but much more powerful
than 
graph-based 
mapping techniques \cite{tket-placement.dagstuhl.2019}
that search for dependence chains via graph traversals.

The
\ourtool{} mapping algorithm is customized
to the abstractions and reasoning
methods of polyhedral frameworks.
Namely, 
we introduce 
a dependence-driven, layer-based objective function
to make key mapping decisions.
The new abstractions and objective function
lead to 
strong improvements in two key
compilation metrics, the {\em circuit depth}
(critical path of the DAG circuit representation)
and {\em SWAP count}, 
which
are critical because qubit states decohere
exponentially in time. Longer circuits yield
higher error rates when measuring.

To summarize, the contributions of this paper are:
i) The \ourtool{} mapping algorithm, the first qubit mapping
and SWAP-insertion optimization algorithm based on transitive
dependences with a global view.
ii) A prototype implementation 
of \ourtool{}, which will be made avaialble.
iii) A thorough evaluation
comparing our approach with four
established baselines, on current back-ends
and future large-scale
synthetic QPUs, and
on QUEKO \cite{queko} and QASMBench \cite{qasmbench} circuits,
demonstrating the advantages
of our algorithm driven by
transitive dependences.

\begin{figure}[h!tb]
  \centering
  \includegraphics[width=\linewidth, trim=5 5 5 5, clip]{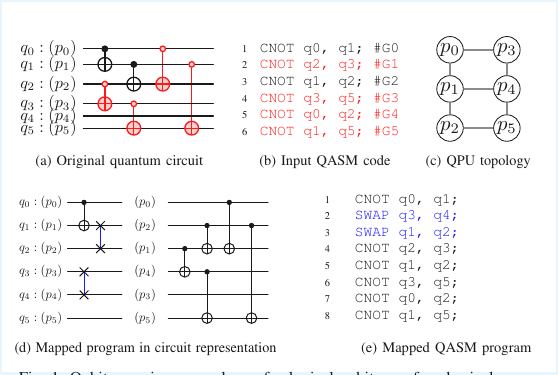}
  \caption{Qubit mapping example. $q_i$ are logical qubits, $p_j$ are physical ones.}
  \label{fig:orig-circuit}
\end{figure}

\section{Motivation}
\label{sec:motiv}
{\bf Qubit Mapping.}
Let us take the example of a quantum program, expressed as a circuit (Fig.~\ref{fig:orig-circuit}a)
or in its QASM representation (Fig.~\ref{fig:orig-circuit}b),
a coupling graph $\mathcal{C}$ (Fig.~\ref{fig:orig-circuit}c)
defining the pairs of directly connected
physical qubits.
In Fig.~\ref{fig:orig-circuit}b, line 2, the CNOT gate reads the logical qubits $q_2$ and $q_3$, which are mapped to the physical qubits $p_2$ and $p_3$, not physically connected. SWAP gates are needed in this case to exchange the quantum states of $p_2$ and $p_3$.
In general, the task of \emph{Qubit Mapping} consists in deciding the placement of qubits and
inserting SWAP gates when needed.
For every two-qubit gate $g_i (q_a, q_b)$, and assuming that $q_a$ and $q_b$ are mapped to $p_a$ and $p_b$, if the pair $\langle p_a, p_b \rangle \not\in \mathcal{C}$, then one or more SWAP gates must be added before $g_i$ (See Fig.~\ref{fig:orig-circuit}d--\ref{fig:orig-circuit}e).
Notably, a SWAP can only be applied to directly connected physical
qubits.

{\bf Complexity of the Qubit Mapping Problem.}
Finding an
optimal qubit mapping for \qc{} programs is exponentially
complex~\cite{siraichi.cgo.2018}.
This complexity is rooted in two factors.
First, rearranging any configuration of n qubits can require up to $O(n^2)$ SWAP operations, making the complexity increase rapidly as the number of qubits grows~\cite{token-swapping.tcs.2015,token-swapping-complexity.algorithmica.2018}.
Second, inserted SWAP gates can introduce new
disconnections in the circuit, further exacerbating the
problem.
As a result, most graph-based mapping strategies
use some variation of the all-pairs-shortest path
algorithm.

\noindent
 \begin{table}[htb]
 \TableFontSize
 \setlength{\tabcolsep}{3pt}
 \vspace{-3ex}
 \caption{Closely related work on quantum mapping}
 \vspace{-1ex}
 \label{tab:motiv}
 \centering
 \begin{tabular}{p{1.2cm}|p{1.0cm}|p{1.95cm}|p{3.5cm}}
 \toprule
 {\bf Work} &
 {\bf Approach} &
 {\bf \# Layers} &
 {\bf Cost Function}
 \\

 \toprule
 
 QMAP\cite{qmap-optimal-subarch-qmap.tqc.2023} &
 DAG   &
 Multi &
 A*-search
 \\

 Sabre\cite{sabre} &
 DAG &
 2 & 
 Decay + Qubit Distance
 \\

 Cirq\cite{circ} &
 DAG & 
 Time-Sliced &
 Qubit Distance
 \\

 tket\cite{tket} &
 DAG &
 Time-Sliced &
 Bounded longest distance
 \\

 {\bf Our work} &
 Affine abstractions &
 Multi-layer, Dependence Distance &
 Layers normalized by transitive dependence volume  
 \\

 \bottomrule
 \end{tabular}
 \shrink
 \end{table}

 {\bf Prior Work.} 
 We briefly summarize the most closely related
 prior work on {\em Qubit Mapping}.
 QMAP \cite{qmap-optimal-subarch-qmap.tqc.2023}
 a mapper from the Munich Quantum Toolkit (MQT),
 encompasses exact and heuristic-based
 mapping strategies. 
 In our evaluation, we leverage
 their heuristic 
 cost function based on A*-search, which divides quantum circuits into layers,
 performs optimal decisions within layers,
 followed by reconciliation passes
 to fix inconsistent mappings between layers \cite{jku.date.2018,wille.dac.2019,wille2023mqt}.
 Unlike QMAP, Sabre \cite{sabre} only splits circuits between a front and an extended layer. 
 Their efficiency
 is based on a heuristic cost
 that accounts for qubit decay while
 assigning distinct weights to the front and
 extended layer. Notably, our approach
 can be seen as a hybrid between these mappers,
 by taking into consideration multiple layers and assigning a different weight to each of them. 
 The key difference is the manner in which 
 we form layers (by dependence distance) and how the layer-weights are computed (volume of transitive dependences). 
 Finally, Cirq and tket use variations of physical
 qubit distance that aim to minimize or bound
 the maximal distance between qubits.

 {\bf Advantages of a Dependence-Driven Affine Framework.}
 The qubit mapping algorithm introduced
 in this work, \ourtool{}, leverages
 the power of mature polyhedral frameworks,
 ISL \cite{isl} and Barvinok \cite{sven.barvinok.2007}, to count
 integer points bounded by polyhedra.
 Once a quantum program is lifted
 from its QASM representation into
 polyhedral abstractions, the analysis
 shifts from individual gates to
 macro-gates (groups of gates that
 follow the same affine behavior and overall structure).
 The key insight of this work is that
 the effect of SWAP gate insertions
 over later circuit regions can be 
 computed via {\bf transitive dependences}.
 

Counting transitive dependences is a crucial ability
since it permits \ourtool{} to make decisions
that later affect fewer qubits.
That is to say, transitive dependences
enable us to insert SWAPs while carefully
deciding the qubits that should be moved
so as to minimize later disruptions in quantum circuit.

\begin{figure}[h!]
\shrink[2]
\centering
\includegraphics[width=0.9\linewidth]{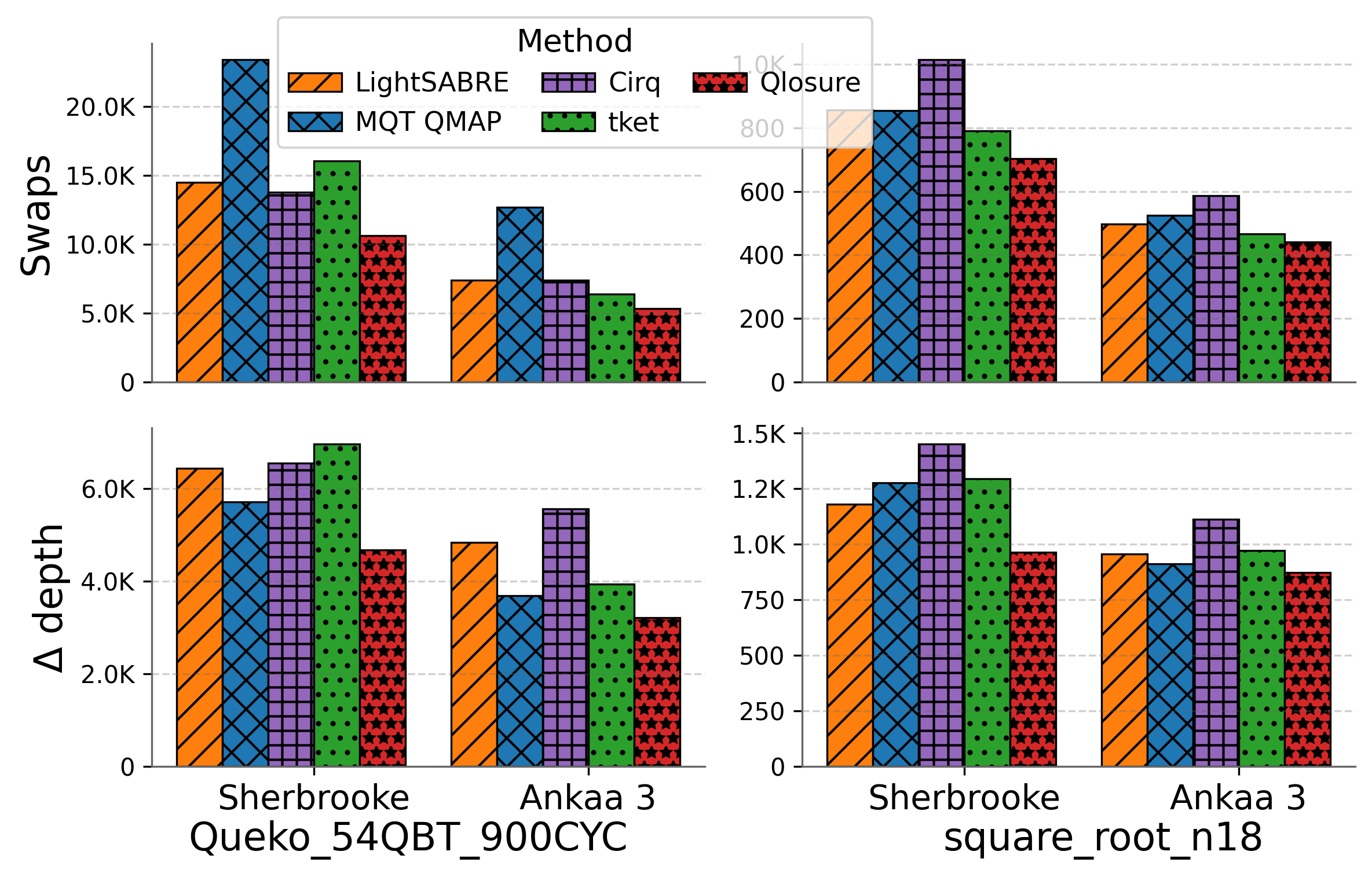} 
\vspace{-1ex}
\caption{
Mapper comparison: Two circuits are mapped onto the IBM Sherboooke and Rigetti Ankaa-3 QPUs.
(i) a 54-qubit circuit (initial depth 900, 9720 two-qubit); and 
(ii) an 18-qubit circuit (initial depth 1429, 898 two-qubit gates).
Metrics of interest are the $\Delta$ (final depth - initial depth) 
and the SWAP count.
}
\label{fig:motivation_plot}
\shrink[1]
\end{figure}

{\bf Mapping Impact.}
We show the impact of \ourtool{} in Fig.~\ref{fig:motivation_plot}, an 
excerpt of our results,
which shows a circuit from the QUEKO
benchmark suite 
\cite{queko} and
one from QASMBench \cite{qasmbench}.
Both circuits are evaluated on two QPU,
IBM Sherbrooke and Rigetti Ankaa-3.
Results of our dependence-driven
mapping are contrasted against 
established mappers: LightSabre \cite{zou2024lightsabre}
(a faster version of Sabre \cite{sabre}),
QMAP \cite{qmap-optimal-subarch-qmap.tqc.2023},
Quantinuum's TKET \cite{tket-placement.dagstuhl.2019,tket},
and Google's Cirq \cite{circ}.

\vspace{-1ex}
\noindent

\section{Background}
\label{sec:background}

\subsection{Polyhedral Model Fundamentals}

The polyhedral model~\cite{Feautrier2011} is a mathematical model for representing code and code transformations and is used in state-of-the-art compilers to apply complex code transformations and reason about their correctness~\cite{Iri88,feautrier_array_1988,wolf1991loop,lefebvre_automatic_1998,Qui00,thies_unified_2001,Darte_contraction_2005,bondhugula_practical_2008,10.1145/3410463.3414635, 10.1145/3581784.3607096, Chatarasi2016AnEP, KONG2013POLYHEDRAL, STOCK2014, baghdadi2015PhD,baghdadi2019tiramisu,baghdadi2018tiramisu1,trifunovic_graphite_2010,polly,tobias_hexagonal_cgo13,Vasilache2018TensorCF,baghdadi2011speculation,merouani2020deep,pouchet.11.popl,baghdadi2020tiramisuDNNDenseSparse}.


\subsubsection{Integer Sets and Relations}
An \emph{integer set} is a collection of integer points defined by a Presburger
formula. 
For example, the integer set 
{\FormulaSize $S = \{\, (i, j) \in \mathbb{Z}^2 \mid 1 \leq i \leq 2, 1 \leq j \leq 2 \}$}
consists of the points {\FormulaSize $\langle i,j \rangle$:
$(1,1),\ (1,2),\ (2,1),\ (2,2)$}.
Similarly, 
an \emph{integer map} defines a relationship between an input and output sets specified by a Presburger formula. 
For example, the following map shifts every point
{\FormulaSize $p \in S$} by 2:
{\FormulaSize $R = \{\, (i, j) \rightarrow (i', j') \mid i' = i + 2,\ j' = j + 2,\ (i, j) \in S \,\}$}.

\subsubsection{Program Representation}
We use the following simple loop to describe
key affine abstractions:

\begin{minipage}{\linewidth}
\begin{lstlisting}[language=c,basicstyle=\ttfamily\footnotesize] 
for (k = 0; k < 5; k++) 
  A[k+1] = A[k] + 1;  // Statement S
\end{lstlisting}
\end{minipage}

The four key abstractions of affine frameworks
for the above loop are: 
i) iteration domains
group statement instances (gates in this work), modeled as n-dimensional
integer sets;
ii) access relations: map iteration domain points to data space,
represented as integer relations;
iii) program schedule; 
iv) polyhedral dependences: relations among
points in iteration domains of program statements.
In our example, the iteration
domain of statement S is {\FormulaSize $\mathcal{D}^S = \{S[k]: 0 \leq k < 5 \}$}.
The statement consists of two references
to $A$, {\FormulaSize $\{ S[k] \rightarrow A[k+1] \}$} (write-reference) and
{\FormulaSize $\{ S[k] \rightarrow A[k] \}$} (read reference).
The schedule of $S$ is represented via the (identity) relation 
{\FormulaSize $\{S[k] \rightarrow [k]\}$}.
From the previous abstractions, exact polyhedral dependences can be computed. For the example, 
the only dependence is {\FormulaSize $\mathcal{D}^{S \rightarrow S} = \{ S[k] \rightarrow [k+1] : k \in \mathcal{D}^S \}$}. 


{\em Transitive Closure of Relations.}
We use the transitive closure of relations~\cite{verdoolaege-transitive-closure.isas.2011}, denoted as $R^{+}$, to estimate the impact of inserted SWAP gates on the rest of the circuit. In simple terms, the transitive closure of a relation R, represented as a graph, allows us to identify all nodes in that graph that are reachable from a given node.
Notably, we apply the transitive closure on relations that are
{\bf polyhedral dependences mapped on the time space}. More about the polyhedral model in Appendix~\ref{appendix:transitive}.

\shrink[0.5]
\subsection{Quantum Computing Fundamentals}


{\em Qubits} (quantum bits)
are the fundamental unit of quantum information. Unlike a classical bit (which can be 0 or 1), a qubit's state, often denoted as \(\lvert \psi \rangle\), can be \(\lvert 0 \rangle\), \(\lvert 1 \rangle\), or a linear combination (superposition) of the two (more in Appendix~\ref{appendix:qubit}).

{\em Quantum Gates}
are the basic operations in quantum circuits, analogous to logic gates in classical computing. Each gate applies a transformation to one or more qubits.
\emph{Single-qubit gates} act on individual qubits,
while \emph{Multi-qubit (typically 2Q) gates} act on two or more qubits simultaneously. One such example is the SWAP gate, which exchanges the states of two qubits: for instance, applying SWAP to \(\lvert q_1 \rangle \lvert q_2 \rangle\) results in \(\lvert q_2 \rangle \lvert q_1 \rangle\), which is particularly relevant for adapting circuits to hardware constraints.
Fig.~\ref{fig:orig-circuit}d shows two types of two-qubit gates, the 
Controlled-NOT (CNOT or CX) and the SWAP gate.

{\em Quantum Circuits}
consist of a sequence of quantum gates applied to qubits, describing how their quantum states change over time.
A crucial concept is \emph{universality}, which ensures that any circuit can be decomposed into a sequence of single-qubit gates and two-qubit gates \cite{deutsch1995universality}.

{\em Quantum Hardware.}
A \emph{Quantum Processing Unit} (QPU) is the physical chip that hosts qubits and performs quantum gate operations. In the NISQ era, QPUs 
are limited to nearest-neighbor connectivity \cite{preskill2018quantum}, 
which is defined by a \emph{hardware graph} 
(where vertices model physical qubits, and edges 
denote direct connections).
When a gate requires two non-adjacent
qubits, one or more \emph{SWAP} gates must be inserted.

\smallskip
\noindent\textbf{Example:}
Let's consider a linear topology consisting of three qubits arranged as: $q_1 \!-\! q_2 \!-\! q_3$. To apply a CNOT gate with $q_1$ as the control and $q_3$ as the target, we must first make them adjacent. This can be achieved by inserting a SWAP gate between $q_2$ and $q_3$:
$\text{SWAP}(q_2, q_3) \Rightarrow q_1 \!-\! q_3 \!-\! q_2$. Now, $q_1$ and $q_3$ are adjacent, allowing the execution of the $\text{CNOT}(q_1, q_3)$. This illustrates how hardware constraints in NISQ QPUs necessitate routing strategies to enable logical gate execution.

\shrink[0.5]
\subsection{Affine Abstraction via QRANE}
\label{ssec:qrane_extraction}
In our work, we employ the QRANE framework 
\cite{qrane}
to 
lift raw QASM circuits into a structured affine form, enabling precise dependency analysis.
To make an analogy with programs in classical computing, a qubit i is represented as $q[i]$ and individual gates are analogous to statement instances.
QRANE groups quantum gates whose qubit arguments exhibit the same affine access 
relation (constant $\times$ i + constant), raising the granularity of the analysis from individual gates (statement instances) to macro-gates (statements).
This level of representation 
provides a scalable foundation for subsequent polyhedral operations.
Once the quantum circuit has been lifted,
precise dependence analysis can be performed to compute the transitive
dependences needed for qubit mapping. Below we show 
a simple input QASM trace and the resulting polyhedral abstractions representing it.
\\

\noindent
\begin{minipage}{\linewidth}
\FormulaSize
\begin{minipage}{0.33\linewidth}
{\bf QASM trace} \\
{\tt CX q[0],q[1];} \\
{\tt CX q[1],q[3];} \\
{\tt CX q[2],q[5];} \\
{\tt CX q[3],q[7];} 
\end{minipage}
\hfill
\begin{minipage}{0.65\linewidth}
{\bf Lifted Representation} \\
Iteration domain: $S = \{ [i] : 0 \leq i \leq 3 \}$; \\
Qubit relations: $q_1 = \{ [i] \rightarrow [i] \}$; ~~~ $q_2 = \{ [i] \rightarrow [2i+1] \}$; \\
Schedule: $\{ [i] \rightarrow [i] \}$
\end{minipage}
\end{minipage}
\\

\section{Affine Dependences for Circuit Mapping}
\label{sec:locality}


Mapping algorithms frequently rely on a \emph{geometric} score~\cite{sabre,zou2024lightsabre,tket-qubit-routing.dagstuhl.2019} that rewards any swap
that shortens the Euclidean or Manhattan distance between the qubits. Such a
metric is attractive for its $O(1)$ evaluation cost, 
after computing the All-Pairs-Shortest-Paths, but it also ignores how the
gate is embedded in the
\emph{dependence graph} (DG). By focusing only on local distance
reductions, it may choose swaps that adversely affect the overall schedule. In
large circuits, this limitation becomes especially pronounced: it can force the
insertion of many additional swaps and substantially increase circuit depth,
since each swap is selected without considering its effect on subsequent gate
placements.

\subsection{Value of Dependence Information}
Dependency analysis elevates circuit mapping from a local optimization with
distance only into a holistic spatio-temporal optimization. By constructing a
dependence graph (DG) and computing its transitive closure, each gate is
annotated with a count of downstream dependents.  This count serves as a proxy
for the criticality of a gate: gates that block many successors truly define the
depth of the circuit, while gates with few dependents offer low-risk and
high-flexibility opportunities to free resources. With this temporal context, a
mapper can balance immediate locality gains against long-term impact,
prioritizing swaps that accelerate critical gates or cheaply clear qubits for
future use, yielding far fewer redundant operations and substantially lower
overall depth than purely geometric heuristics, all with scalable performance on
large circuits.

\subsection{Dependency Computation via Transitive Closure}
\label{ssec:isl_transitive_closure}

Modeling the spatio-temporal behavior
of circuits demands capturing 
the point-to-point dependence relations
(flow, anti, output, and read)
and the transitive 
dependences. To do this, we leverage the
Integer Set Library (ISL)~\cite{isl}, which 
provides polyhedral operations,
dependence analysis support and \emph{transitive closure} operator:

{\em Dependence Relations} ($R_{dep}$).  
The dependence graph of the input program is computed
from lifted affine abstractions.
{\em Closure Operator.}  Applying the transitive-closure operation to the dependence relations
{\FormulaSize ~ $R^{+} = \text{transitive\_closure}(R_{\text{dep}}) ~ $}
yields the transitive dependency relation, capturing all the successors of a gate in the dependence graph.
        
{\em Dependence Weight.}  
Once \(R^{+}\) is available, 
the transitive successor count 
{\FormulaSize \(\,\omega_{g} = \text{card}(\{\,h \mid (g,h)\in R^{+}\})\)},
offered by the Barvinok library \cite{sven.barvinok.2007},
is computed once.
This enables our analysis to query, in $O(1)$ time, each gate \(g\), providing a precise measure of its impact on downstream operations.
        

\section{Mapping Circuits}
\label{sec:mapping}
We now introduce the \ourtool{} algorithm and cost
function heuristic used to overcome the limitations of prior work.

\shrink[0.5]
\subsection{Overview}

\begin{figure}[tbh]
\shrink[1]
\centering
\includegraphics[width=\linewidth]{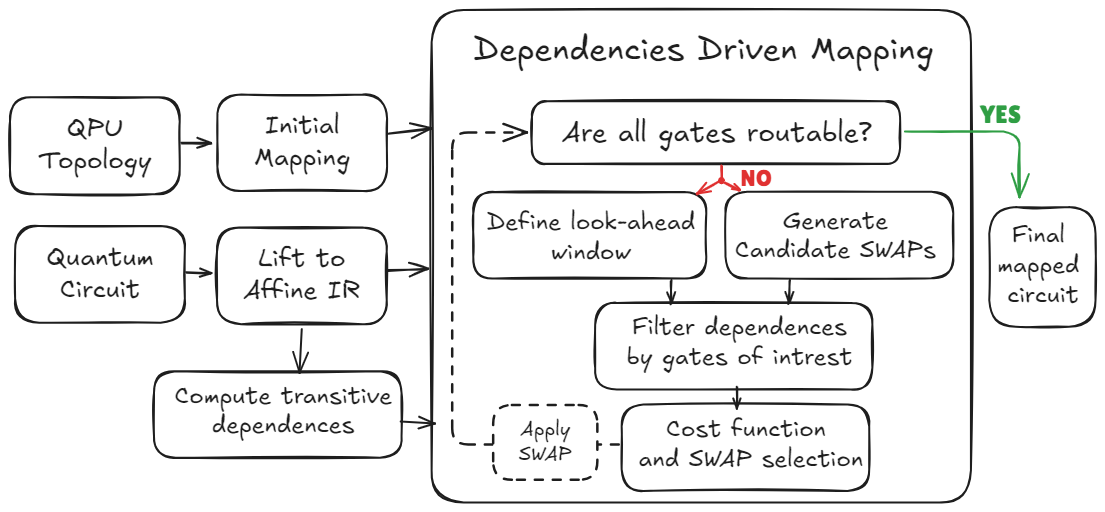}
\caption{Qlosure Mapping Pipeline}
\label{fig:qlosure_pipeline}
\shrink[1]
\end{figure}

\begin{figure}[tbh]
    \shrink
    \centering
    \captionsetup[subfigure]{justification=centering, font=footnotesize, labelformat=parens, labelsep=space}

    \begin{subfigure}[b]{0.48\linewidth}
        \centering
        \includegraphics[width=\linewidth]{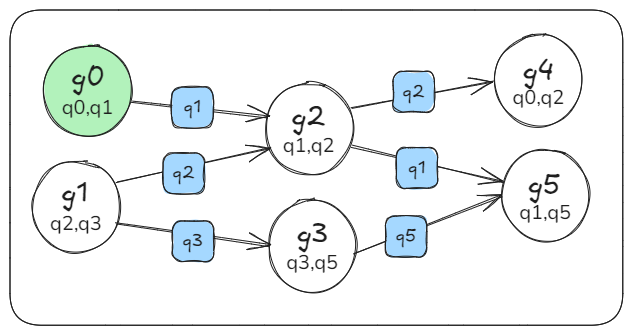}
        \caption{Dependence graph}
        \label{fig:map-a}
    \end{subfigure}
    \hfill
    \begin{subfigure}[b]{0.48\linewidth}
        \centering
        \includegraphics[width=\linewidth]{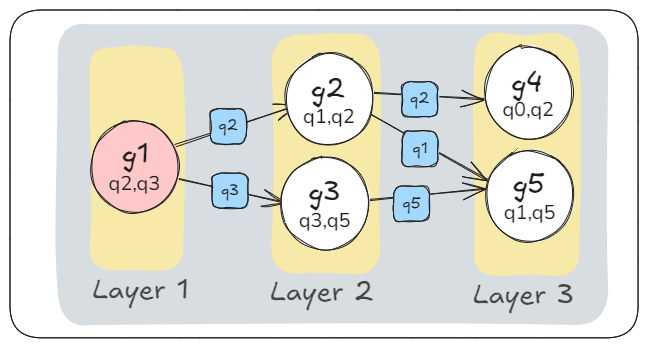}
        \caption{Layered look-ahead window}
        \label{fig:map-b}
    \end{subfigure}

    \vspace{0.2em}

    \begin{subfigure}[b]{0.52\linewidth}
        \centering
        \includegraphics[width=\linewidth]{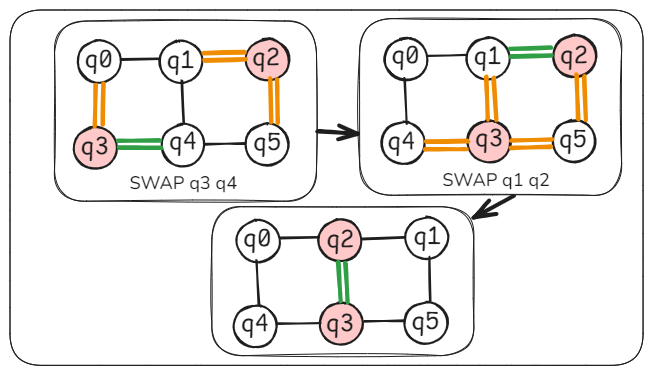}
        \caption{SWAP candidates (orange/green), selected SWAP (green)}
        \label{fig:map-c}
    \end{subfigure}
    \hfill
    \begin{subfigure}[b]{0.42\linewidth}
        \centering
        \includegraphics[width=\linewidth]{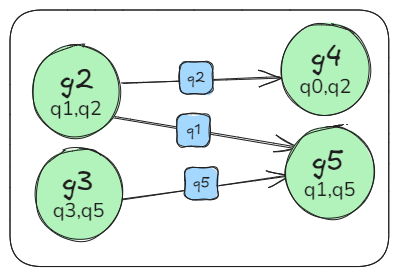}
        \caption{Final state with all gates executable}
        \label{fig:map-d}
    \end{subfigure}

    \caption{Mapping example.}
    \label{fig:mapping-overview}
    \shrink[2]
\end{figure}

Fig.~\ref{fig:qlosure_pipeline} depicts 
the flow of our
qubit mapping algorithm. From an unmapped
quantum circuit and QPU topology, we
first lift the circuit to an \emph{Affine Intermediate
Representation}~\cite{qrane}. 
Each logical qubit is initially assigned to a
distinct physical qubit.
The main part of the algorithm is the \emph{Dependence-Driven Mapping} 
loop,
which we explain via the example in Fig.~\ref{fig:mapping-overview}.

After building the \emph{dependence graph} DG 
(Fig.~\ref{fig:map-a}) 
and computing its \emph{transitive closure}, the
algorithm enters the \emph{Dependence-Driven Mapping} (DDM) loop.
At every iteration we: 
(i) identify the set $L_{\mathrm f}$ of \emph{ready} gates
(gates with no unsatisfied dependencies); 
(ii) test each gate  \(g\in L_{\mathrm f}\) 
for
\emph{hardware feasibility}, that is, 
whether used qubits are adjacent on the
target device under the current logical-to-physical mapping 
\(\varphi : Q_{\text{logical}} \to Q_{\text{phys}}\).
\noindent
If qubits of $g$ are adjacent on the coupling graph,
$g$ is executed and removed from the DG, otherwise a SWAP sequence must be inserted to enable its execution.

Initially, gates $g_0$ and $g_1$ are both ready (in \(L_{\mathrm f}\)), but only
$g_0$ can be executed because its used qubits \((q_0,q_1)\) are already
adjacent; therefore, we execute \(g_0\) and update the DG, removing $g_0$, 
and
recompute the set \(L_{\mathrm f}\).
Now $g_1$ remains the only ready gate (Fig. \ref{fig:map-b}), yet its used qubits
\((q_2,q_3)\) are disconnected on the target QPU
(see Fig.~\ref{fig:orig-circuit}c).
To resolve this, we generate candidate SWAPs, as described in
Sec.~\ref{sub-sec:candidate-swap-generation}. 
The candidate SWAPs are
illustrated in Fig.~\ref{fig:map-c}, with green and orange edges. Each SWAP is
evaluated using the heuristic cost function \HeuristicCost{}
(Sec.~\ref{sec:swap-cost}), which estimates the SWAP impact on 
current and upcoming gates. The SWAP with the lowest score is selected and
applied, updating the mapping \(\varphi\).

If \(g_1\) remains infeasible after the SWAP (see Fig~\ref{fig:map-c}), the
process continues, applying additional SWAPs one at a time until \(g_1\) or
another gate in \(L_{\mathrm f}\) becomes executable. Once a gate executes, it is removed
from the DG and the set \(L_{\mathrm f}\) is updated accordingly. The DDM loop then
proceeds with the next iteration.
We remark that SWAP decisions are made from gates in the
{\bf look-ahead window} ($L_{\mathrm w}$) whose impact over
the entire circuit is computed via transitive dependences.

The algorithm terminates when the DG becomes empty, meaning all gates have been scheduled and executed in compliance with the hardware connectivity (Fig.~\ref{fig:map-d}). The resulting mapped circuit requires no additional SWAPs.

\subsection{\label{sec:initialization}Initialization and Precomputation}
We now introduce definitions and intermediate
structures used to map qubits.
Broadly, these constructs allow us
to manipulate polyhedral dependence abstractions,
compute all transitive dependencies, 
identify affected qubits, and separate
quantum gates into layers for easier handling.

\subsubsection{New Polyhedral Abstractions}

In addition to the abstractions already 
discussed in Sec.~\ref{sec:background},
we introduce three new abstractions used to map quantum circuits.
These abstracts are expressed using
{\em integer sets} and {\em maps}.

{\em Hardware connectivity.}
We model the physical connectivity of qubits
with the set $R_{hw}$:

\noindent
\begin{minipage}{\linewidth}
\begin{footnotesize}
\begin{align*}
      R_{\mathrm{hw}} = \{(p1,p2) \mid\ &p1,p2 \in Q_{phys},
      \text{$p1$ and $p2$ are adjacent physical qubits} \}
\end{align*}
\end{footnotesize}
\end{minipage}
\vspace{0.25ex}

\noindent
where \( Q_{phys} \) denotes the set of \emph{physical qubits} available on the quantum processor.
The above set captures the device’s qubit topology 
as a set of pairs. If two physical qubits can directly interact, the pair is included in the set.

{\em Use Map.}
To facilitate gate-to-qubit association, we  
introduce the {\em Use Map} abstraction, which maps  
each logical multi-dimensional time-step, 
$\vec{t} \in \mathbb{T}$, to the pair of logical qubits involved in the gate executed at that step. 
Here, \( \mathbb{T} \) denotes the set of all time steps in the circuit's logical schedule.
Specifically, the map  $U \colon \mathbb{T} \to Q_{logical} \times Q_{logical}$
returns the pair of logical qubits $(q_1, q_2)$ on which the gate at time $\vec{t}$ operates,  where \( Q_{logical}\) denotes the set of logical qubits in the circuit.

 %

\noindent
{\em Dependence map:}
Prior to computing the dependence transitive closure, we compute
all the distinct point-to-point dependences in the program -- excluding
possible transitive dependences. 
Let $
\mathcal{O} \;=\; \{(\vec{t},  q_1, q_2)\mid \vec{t}\in\mathbb{T},\;(q_1, q_2) = U(\vec{t})\}
$
be the set of all gate instances along with the qubits they access.  We then define 

\vspace{-1ex}
\noindent
\begin{minipage}{\linewidth}
\begin{footnotesize}
\begin{eqnarray*}
    R_{\mathrm{dep}}
    =
    \bigl\{
      [(\vec{t_1},q_1,q_2)]\to[(\vec{t_2},q_3,q_4)] : 
    \quad (\vec{t_1},q_1,q_2) \in \mathcal{O},
    \\
    (\vec{t_2},q_3,q_4) \in\mathcal{O},
    \quad \vec{t_1} \prec \vec{t_2},
    \quad q_1 \neq q_2 , \quad q_3 \neq q_4, 
    \\
    \quad \bigl(q_1 = q_3 \;\vee\; q_1 = q_4 \;\vee\; q_2 = q_3 \;\vee\; q_2 = q_4\bigr) \Bigr\}.
\end{eqnarray*}
\end{footnotesize}
\end{minipage}
\vspace{0.5ex}

\noindent
which captures all \emph{flow}, \emph{anti}, \emph{output} and \emph{read}
dependencies whenever two gates share a logical qubit.
\subsubsection{Dependence Weight Function ($\texorpdfstring{\omega}{omega	extunderscore g}$)}  To prioritize swaps that affect the least critical gates, we count, for each gate $g$, the number of dependents (i.e., successors) it has in the transitive‐closure of the dependence map.
$\omega_g$
is used
to compute the $\HeuristicCost$-score (defined later). 
It assigns to each gate instance \(g=(\vec{t}, q_1, q_2)\) the number of its transitive dependents (a gate instance is uniquely identified by the time-step $\vec{t}$):

\noindent
\begin{minipage}{\linewidth}
\begin{footnotesize}
\begin{equation}
\begin{split}
\hspace{-4ex}
\omega\colon \mathcal{O} \to \mathbb{N}, 
\hspace{0.67\linewidth}
\\
 \omega(\vec{t},q_1, q_2) = \operatorname{card}\left( 
 \left\{ 
 (\vec{t}', q_1', q_2') \mid
 (\vec{t}, q_1, q_2) \rightarrow (\vec{t}', q_1', q_2') \in R^{+} 
 \right\} 
 \right)
 \\
\end{split}
\label{eq:omega}
\end{equation}
\end{footnotesize}
\end{minipage}
Later in the paper, we use $w_g$ as shorthand for $\omega(\vec{t},q_1, q_2)$.

Eq.~\eqref{eq:omega} will later be used to assign weights
to gates in the {\bf Front Layer} ($L_{\mathrm f}$)
and {\bf Look-ahead Window} ($L_{\mathrm w}$) proportional
to the number of downstream dependent operations.
$R^{+}$ in Eq.~\eqref{eq:omega} 
is computed once using the transitive 
closure 
of the dependence relations, via ISL,
and captures all direct and transitive dependence routes.

\subsubsection{Distance matrix\label{sec:distancematrix}}
Given the hardware graph $(Q_{phys}, R_{\mathrm{hw}})$,  
we compute the All-Pairs Shortest Paths 
between physical qubits 
\cite{wille.aspdac.2014}.
The result is a symmetric \(|Q_{\mathrm{phys}}| \times |Q_{\mathrm{phys}}|\) matrix \(D_{\mathrm{phys}}\), where each entry \(D_{\mathrm{phys}}[p_1, p_2] \in \mathbb{N}\) gives the minimum number of SWAPs  
needed to move a qubit from $p_1$ to $p_2$ reflecting the physical separation between them.

\subsubsection{Initial mapping}
Before the swap‐based mapping algorithm can begin, we 
fix an initial assignment of logical qubits to physical qubits \cite{qubit-init-placement-influence.2019}.  
Unless otherwise noted, we use the identity mapping (i.e., logical qubit \(q_i\) is placed on physical qubit \(p_i\)):
{\FormulaSize $\phi_0: Q_{logical} \;\longrightarrow\; Q_{phys}, ~ \phi_0(q_i) = p_i$}.  
We also note that the qubit mapping obtained by our mapping algorithm can be
used as the initial qubit assignment for a second run of the algorithm,
similar to
\cite{sabre},
leading to even better results. This option is explored in the
ablation study of Sec.~\ref{sec:eval:ablation}.


\vspace{-0.5ex}
\subsection{Layers and Look-Ahead Window}
\label{sec:layers}
Throughout the algorithm, we maintain a dynamic structure 
to manage gate
scheduling and SWAP decisions. The structure includes the \emph{Front Layer}
($L_{\mathrm f}$), which consists of all gate instances for which all required
preceding gates in $R_{\mathrm{dep}}$ have already been executed.
While $L_{\mathrm f}$ remains  non-empty we dequeue and execute every gate whose two mapped qubits are adjacent in $R_{\mathrm{hw}}$,
updating \(L_{\mathrm{f}}\) accordingly until no further gate is executable without inserting a SWAP.

To enhance SWAP selections, a \textbf{look-ahead window} $L_{\mathrm w}$ is
maintained. This window extends beyond $L_{\mathrm f}$ to include sets of gates that will
soon become executable and that act on physical qubits currently used
by gates in $L_{\mathrm{f}}$. Gates in $L_{\mathrm w}$
are considered likely to
\emph{affect or constrain} near-future scheduling if not taken into account. The
size $k$ of the look-ahead window is dynamic: $k \;=\; c \, n_{\mathrm f}$,
where $n_{\mathrm f} = |\text{qubits}(L_{\mathrm f})|$ denotes the number of
distinct physical qubits involved in the front layer. The constant $c$ is chosen
based on the hardware topology, specifically, it is set to exceed the maximum
degree of any node in the hardware coupling graph $R_{\mathrm{hw}}$, ensuring
the window is wide enough to capture potential interactions around the active
region. This window \(L_{\mathrm w}\) is formed by selecting the topologically
earliest \(k\) gate instances from the dependency map \(R_{\mathrm{dep}}\).

Gates in $L_{\mathrm w}$ are further organized into layers \(\mathcal{G} = \{G_1, G_2, \dots\}\),
where $G_1 = L_{\mathrm f}$ and each subsequent layer $G_{i+1}$ includes gates
that become executable only after all gates in $G_i$ have been executed.
These layers are defined based on the \emph{dependence distance} \(\ell\) in the
dependency map \(R_{\mathrm{dep}}\), capturing how far each gate is from the
current front in terms of dependency levels.
The union of all layers covers the look-ahead window, i.e., \(\bigcup_i G_i = L_{\mathrm w}\). 
This layered view captures both immediate and emerging dependencies, providing a
temporal-aware view for more strategic routing decisions. Gates in earlier layers are closer to execution and thus more important for routing, as their qubits should be made adjacent first. Deeper-layer gates represent future operations and are considered with reduced influence, contributing proportionally less to swap decisions.


    
\subsection{Candidate-SWAP generation}\label{sub-sec:candidate-swap-generation}
We define $\mathcal N(p)$ as the set of physical neighbors of
qubit $p$ in $R_{\mathrm{hw}}$ and  $P_{\mathrm{front}}$ as the set of physical qubits currently mapped to logical qubits that participate in gates from the front layer $L_{\mathrm{f}}$.
For each $p_1 \in P_{\mathrm{front}}$, we consider all possible SWAP operations between $p_1$ and each of its neighbors $p_2 \in \mathcal{N}(p_1)$.  
The set of candidate SWAPs is then:

\vspace{1ex}
\begin{minipage}{\linewidth}
\begin{footnotesize}
\begin{equation*}
\mathcal{S} = \bigcup_{p_1 \in P_{\mathrm{front}}} \{ (p_1, p_2) \mid p_2 \in \mathcal{N}(p_1) \},
\end{equation*}    
\end{footnotesize}
\end{minipage}

\noindent
Each candidate~$s\in\mathcal S$ is evaluated by the cost
function~$\HeuristicCost$ defined in Sec. \ref{sec:swap-cost}
and the swap with minimum cost is applied.

\subsection{Qlosure SWAP-cost heuristic}
\label{sec:swap-cost}
We now introduce the {\bf cost function} used
to make swap insertions and qubit mapping 
decisions.
Given the current logical‐to‐physical assignment
$\varphi\colon Q_{logical}\!\to\!Q_{phys}$,  
let $s=(p_1,p_2)\in R_{\mathrm{hw}}$ be a candidate SWAP acting on
the \emph{physical} qubits that presently host the logical qubits
$q_1=\varphi^{-1}(p_1)$ and $q_2=\varphi^{-1}(p_2)$.
Applying $s$ yields a tentative mapping
\[
  \varphi_s
  \;=\;
  \varphi\circ s
  \quad\bigl(
    \text{i.e.\ } \varphi_s(q_1)=p_2,\;
                 \varphi_s(q_2)=p_1
  \bigr).
\]

For all candidate SWAP gates $\mathcal S$, we use a composite function $\HeuristicCost(s)$ to estimate their overall circuit impact:

\vspace{-1ex}
\noindent
\begin{minipage}{\linewidth}
\begin{footnotesize}
\begin{equation}
\begin{split}
  \HeuristicCost(s)
  =
  \max~\!\bigl(\delta_{p_1},\delta_{p_2}\bigr)
  \times
  \left(
  \sum_{G_\ell \in \mathcal{G}}
  \frac{\Gamma_l }{\lvert G_\ell\rvert}
  \right)
  \\
  \Gamma_l = 
  \sum_{g\in G_\ell}
  \frac{\omega_g ~ D_{\mathrm{phys}}\!\bigl[\varphi_s[g_{q_1}],\;\varphi_s[g_{q_2}]\bigr]}
       {\ell}
\end{split}
\label{eq:H}
\end{equation}
\end{footnotesize}
\end{minipage}

The main components of Eq.~\ref{eq:H} are the {\em decay factor} \cite{sabre} (detailed later),
and the swap cost $\Gamma_{\ell}$ assigned to each circuit layer at the current
time-step. 
$G_\ell$ is the set of gate instances whose \emph{level} (distance from $L_{\mathrm f}$)
equals~$\ell - 1$,
as introduced in Sec.~\ref{sec:layers}.
We also clarify that 
$g \in G_{\ell}$ is an arbitrary two-qubit gate with qubit operands $g_{q_1}$ and $g_{q_2}$.
{\bf $\omega_g$} is the computed dependency weight of gate~$g$
(cf.\ Eq.~\eqref{eq:omega}), while $D_{\mathrm{phys}}$ is the distance matrix defined in Sec.~\ref{sec:initialization} and $\ell$ is the topological depth of the layer.
    
Finally, the qubit 
\emph{decay factor} $\delta_{q}$ (initialized to 1) 
discourages  repeatedly swapping the same logical qubit in short
succession \cite{sabre}.  
After committing the best swap, $s^\star$, we  update
  
  {
  \shrink
  \FormulaSize
\[
  \delta_{q} \;:=\;
  \begin{cases}
    \delta_{q}+0.001, & q\in\{q_1,q_2\}\\[2pt]
    \delta_{q},   & \text{otherwise}
  \end{cases}
\qquad
\]
  }
  Once a two-qubit gate is successfully scheduled for execution, $\delta_q$ is reset to its initial value of $1$.

  


{\bf Selection rule.}
The introduced metrics are
used to decide among all candidates
$s\in\mathcal S$ (cf.\ Sec.~\ref{sec:layers}).
We select:

{
\shrink
\FormulaSize
\[
  s^\star \;=\; \arg\min_{s\in\mathcal S} \HeuristicCost(s)
\]
}

\noindent
breaking ties randomly.
The mapping $\varphi$ is updated 
to 
$\varphi \circ s^{*}$ (via map composition),
$\delta$ is refreshed as above, and the algorithm
continues with gate scheduling (for execution) or further SWAP generation
until the circuit completes.


\vspace{-1ex}
\subsection{Intuition}

The objective in qubit routing is \emph{not} just to bring the next
pair of front-layer qubits together.  
Instead, each SWAP should help move the circuit forward as a whole, reducing the
total number of SWAPs and the final depth of the circuit. Each part of the cost
function, Eq.~\eqref{eq:H}, 
drives the algorithm toward better SWAP choices. 
We next provide an intuition
of how each factor contributes to the cost function.

The \textbf{physical distance} between two
qubits is the main metric for local decisions,
i.e., with a window size of 1.
Qubit pairs farther apart on 
the QPU graph
need more SWAPs to interact. 
Hence, swaps that reduce the distance between qubit pairs appearing together in upcoming gates are preferred.

The 
\textbf{dependency weight, $\boldsymbol{\omega_g}$,}
is the critical factor when computing $\HeuristicCost(s)$
across all of $L_{\mathrm w}$.
Gates with fewer future-dependent operations are prioritized because they use qubits that are unlikely to be needed again later in the circuit. Executing such gates early frees up those qubits with minimal interference to the placement of upcoming gates.
Put 
differently, 
a \emph{lower cumulative dependency cost} is better, as it
reflects 
preferring
gates with limited interaction with the 
remaining
circuit, those whose early execution help unlock qubits 
unlikely to be
reused. This reduces scheduling complexity and enables swaps that have minimal
impact on the placement of future gates.


\textbf{Layer discount ($1/\ell$).}  
Within the look-ahead window, each gate at layer $\ell$ (its topological depth) is weighted by $1/\ell$, giving higher priority to gates closer to the front layer. This discounting ensures that deeper gates, whose scores become increasingly small, have reduced influence on the total score, effectively focusing the optimization on imminent operations which reflects the decreasing reliability of predictions about the system's topology further into the future.

\textbf{Layer normalization factor ($1/|G_\ell|$).}  
To ensure that all layers — even narrow ones — contribute comparably, we normalize by the number of gates in each layer. This prevents wide layers from dominating the score and keeps the metric scale-invariant.

For example, if \( G_1 \) contains two gates with scores \( s_1 \) and \( s_2 \), and \( G_2 \) contains three gates with scores \( s_3 \), \( s_4 \), and \( s_5 \), the normalized contributions are:
{
\FormulaSize
$
G_1: \frac{s_1 + s_2}{2}, \quad 
G_2: \frac{s_3 + s_4 + s_5}{3}
$
}


\textbf{Decay factor, $\max(\delta_{q_1},\delta_{q_2})$}.  
Repeatedly swapping the same logical qubit can cause
“thrashing” that bloats depth; the exponential decay
dampens the score of swaps that move \emph{new} qubits while
penalising those that undo recent moves.

The designed 
\HeuristicCost{} function results in fewer total SWAPs and shorter depth
than other mapping methods, as shown in Sec.~\ref{sec:eval}.

\begin{algorithm}[thb]
  \FormulaSize
  \caption{Qlosure qubit mapping and routing}
  \label{alg:qlosure}
  \begin{algorithmic}[1]
    \Require Circuit dependence map $R_{\mathrm{dep}}$, hardware map $R_{\mathrm{hw}}$,
            initial assignment $\varphi:Q_{logical}\!\to\!Q_{phys}$, distance matrix $D_{phys}$,
    \Ensure Fully routed circuit with updated assignment $\varphi$
    \State $L_{\mathrm f} \gets$ gates whose precedences are satisfied
    \State $R^{+} \gets transitiveClosure(R_{dep})$
    \State $\delta_q \gets 1\;\; \forall q\in Q_{phys}$
    
    \While{$L_{\mathrm f}\neq\varnothing$}
      \State $G_{\text{ready}} \gets
             \textsc{ExtractReadyGates}(L_{\mathrm f},\varphi,R_{\mathrm{hw}})$
      \If{$G_{\text{ready}}\neq\varnothing$}
        \State execute $G_{\text{ready}}$
        \State $L_{\mathrm f} \gets
               (L_{\mathrm f}\setminus G_{\text{ready}})
               \cup
               \textsc{Successors}(G_{\text{run}},R_{\mathrm{dep}})$
        \State $\delta_q \gets 1\;\; \forall q\in Q_{phys}$
      \Else
        \State $L_{\mathrm w} \gets \textsc{MakeLookahead}(L_{\mathrm f}, R_{dep})$
        \State $\mathcal{G} \gets \textsc{BuildLookaheadLayers}(L_{\mathrm w},R_{dep})$
        \State $\mathcal S \gets
               \textsc{GenSwapCandidates}(L_{\mathrm f},R_{\mathrm{hw}})$
        \State $s^\star \gets
               \underset{s\in\mathcal S}{\arg\min}\;\HeuristicCost(s,R^{+},\mathcal{G},D_{phys},\delta)$
        \State $\varphi \gets \varphi\circ s^\star$
        \State execute $s^\star$
        \State $\delta \gets \textsc{UpdateDecay}(\delta,s^\star)$
      \EndIf
    \EndWhile
    \State \Return routed circuit and final mapping $\varphi$
  \end{algorithmic}
\end{algorithm}

\vspace{-2ex}
\subsection{End-to-end Qlosure routing algorithm}
\label{sec:qlosure-algorithm}

The previous sections defined all the ingredients Qlosure needs:
the dependence relations $R_{\mathrm{dep}}$,
the hardware graph $R_{\mathrm{hw}}$,
the distance matrix $D_{phys}$, 
and the cost function $\HeuristicCost(s)$.
We now put these pieces together into a single, self-contained
routing loop.
At all moments, the routing algorithm maintains:

\begin{enumerate}[label=(\alph*)]
\item a {\em front layer} $L_{\mathrm f}$ of gates for which all dependency constraints in $R_{\mathrm{dep}}$ are satisfied
\item a {\em dynamic look-ahead window} $L_{\mathrm w}$, $G_1, G_2, \ldots$, 
of the topologically earliest $k = c\,n_{\mathrm f}$
additional gates that may soon become executable,

\item the current qubit mapping
$\varphi:Q_{logical}\!\to\!Q_{phys}$, and

\item a decay vector $\delta$ that tracks how recently each logical
qubit was swapped.

\end{enumerate}

Given this information, the decision-making process
is as follows:
\textbf{If front-layer gates are executable} under
$\varphi$ (their current mapping),
schedule them for immediate execution (i.e., without requiring a swap), and push their children into
$L_{\mathrm f}$.
\textbf{Otherwise} generate candidate SWAPs that touch qubits in
$L_{\mathrm f}$, score each candidate with the $\HeuristicCost$ score,  
apply the best one, and update the mapping $\varphi$.

This loop repeats until 
all gates are executed.
{\bf Alg.~\ref{alg:qlosure}}  details the needed steps
to map qubits and insert SWAPs.

\section{Evaluation}
\label{sec:eval}


We conduct a robust evaluation with three objectives:
a) Demonstrate our efficiency on circuit depth and SWAP count
relative to mapper baselines;
b) Show competitive scalability;
c) Provide evidence that sources of improvement stem from
exploiting dependence information.

\subsection{Experimental Setup}
\label{sec:eval:setup}

The evaluation is conducted on 
the latest stable version
of four state-of-the-art 
mappers.
Mappings were performed on a 2.40 GHz Intel(R) Xeon(R) E5-2680 CPU with 128 GB RAM.
A timeout of 24 hours was imposed for each mapping method to ensure practical runtime limits.

\subsubsection{Qubit Mappers}
We compare \ourtool{} to four state-of-the-art mappers:
{\tt SABRE} (we use the LighSABRE implementation ~\cite{zou2024lightsabre} in QISKIT~\cite{qiskit}, an improved and currently adopted version of the original algorithm),
\texttt{Pytket}~\cite{sivarajah2020t} 
(Python interface to Quantinuum's \texttt{tket}),
\texttt{Google Cirq}~\cite{Cirq} 
and
\texttt{MQT QMAP}~\cite{wille2023mqt} (Munich Quatum Toolkit Quantum Mapper).
We use the identify initial qubit layout for all mappers.
  

\subsubsection{Benchmark Suites}
We evaluate our mapping approach on two established
benchmark suites: 
{\bf QASMBench}~\cite{qasmbench} and {\bf QUEKO}~\cite{tc20-tan-cong-optimality-layout-queko}.
QASMBench 
offers multiple datasets and circuit sizes while
QUEKO, the QUantum Examples with Known Optimal (Depth) suite,
offers
circuits designed
with
provable optimal depth for a given device topology. 
QUEKO circuits are used to 
gauge and estimate the mapping impact of possible
future, denser circuits, while still being substantially
sparse.
We use the methodology described by Tan et al.~\cite{tc20-tan-cong-optimality-layout-queko} to generate 
circuits customized to large, next
generation QPUs. We further discuss this in
Section~\ref{sec:custom_queko_generation}.

\subsubsection{Quantum Hardware Backends}
\label{eval:backends_intorduction}
1) \textbf{IBM Sherbrooke}\cite{ibm_sherbrooke}, a 127-qubit superconducting quantum processor leveraging a heavy-hexagon lattice topology, in which each qubit connects to at most three neighbors in a repeated hexagonal pattern to balance connectivity and error suppression.
2) \textbf{Rigetti ankaa\_3}\cite{rigetti_qpus} is an 82-qubit superconducting processor featuring a modular, multi-chip architecture of square-lattice tiles linked by tunable couplers, enabling scalable two-qubit interactions both within and between chips.
3) \textbf{Sherbrooke-2X} is a synthetic 256-qubit backend constructed by concatenating two IBM Sherbrooke topologies and adding two bridging qubits to form an extended heavy-hexagon lattice. This larger architecture represents a possible evolution of IBM quantum processors, allowing us to evaluate scalability and routing strategies on extended qubit topologies.
Coupling graphs for these backends are shown in Appendix~\ref{appendix:backends}.

\subsubsection{Custom QUEKO Benchmark Generation}
\label{sec:custom_queko_generation}
We use the QUEKO methodology to generate two
new sets of benchmarks with known optimal depth.
The new benchmark sets are synthesized for
an 81-qubit 9x9 and 
a 256-qubit 16x16
grid-shape QPUs.
In these architectures, each internal qubit is designed to connect to its eight nearest neighbors: top, bottom, left, right, and the four diagonal neighbors (top-left, top-right, bottom-left, bottom-right). Qubits located at the edges and corners of the grid have correspondingly fewer connections.

We generate new QUEKO circuits to
stress, 
to different degrees, the physical
connectivity of qubits in the real QPUs back-ends.
Circuits generated from the
81-qubit grid are evaluated on
the Rigetti
Ankaa-3 (an 82-qubit QPU with a maximum of 4 neighbors per qubit) 
and on IBM Sherbrooke (a 127-qubit QPU
with a maximum of 3 neighbors per qubit).
Circuits generated 
from the 16x16 grid are 
 evaluated on 
the synthetic Sherbrooke-2X QPU. 
Each new benchmark set consists of
9 circuit depths and 10 circuits per depth.


\subsection{Summary of Results}
\label{sec:eval:summary}

Tables \ref{tab:depth_summary}--\ref{tab:swaps_summary} 
provide a high‐level summary of Qlosure’s performance against baseline mappers
on the QUEKO benchmark suite, using
circuits generated for 16, 54, and 81
qubits on all three back-ends.
For analysis, we group benchmarks 
by
their initial pre‐mapping circuit depth: 'Medium' (depth $\leq$ 
than 500), and 'Large' (depth $\geq$ 600).

Table \ref{tab:depth_summary} details the average depth‐factor (post-mapping
depth relative to the initial depth, lower is better). 
Across both, Sherbrooke and Ankaa-3 backends. Qlosure consistently
yields the lowest depth‐factors. For Medium‐depth Sherbrooke circuits, Qlosure
achieves a depth‐factor of 5.72, outperforming SABRE (7.68) and Pytket (9.99).
On Large Sherbrooke circuits, Qlosure reaches 5.45 versus 7.18 for SABRE and
9.03 for Pytket. Similarly, on the Ankaa-3 device, 
Qlosure achieves depth-factors
of 4.41 (Medium) and 4.08 (Large), compared to 6.00/5.46 for SABRE and 6.47/5.89
for Pytket. 
On Sherbrooke-2X backend, Qlosure attains depth‐factors of 14.94
(Medium) and 13.45 (Large), while SABRE and Pytket deliver 28.16/24.42 and
37.21/30.93, respectively; QMAP was unable to complete mapping on Sherbrooke-2X
(timeout).

Tab.~\ref{tab:swaps_summary} shifts focus to SWAP efficiency, presenting the average number of SWAPs for existing methods normalized by the number of SWAPs obtained by Qlosure; ratios above 1.0 indicate that existing mapping methods insert more SWAPs than Qlosure. All existing methods, on all backends, without exception, generate more SWAPs compared to Qlosure. For example, on Sherbrooke, SABRE and QMAP generate $1.17\times$ and $1.81\times$ more SWAPs for Medium circuits, and $1.20\times$ and $1.85\times$ more SWAPs for Large circuits.
On the Sherbrooke-2X backend, QMAP again timed out on this larger device.
\noindent
\begin{table}[ht]
  \TableFontSize
  \centering
  \caption{Summary of QUEKO benchmark results: average depth‐factor (post-mapping depth / optimal depth). Lower is better.
  }
  \vspace{-1ex}
  \label{tab:depth_summary}
  \begin{tabular}{p{0.8cm} c c c c c c}
    \toprule
    \multirow{2}{*}{\textbf{Mapper}}
      & \multicolumn{2}{c}{\textbf{Sherbrooke}} 
      & \multicolumn{2}{c}{\textbf{Ankaa‑3}}
      & \multicolumn{2}{c}{\textbf{Sherbrooke-2X}}\\
    \cmidrule(lr){2-3} \cmidrule(lr){4-5} \cmidrule(lr){6-7} 
      & \textbf{Medium} & \textbf{Large} & \textbf{Medium}
      & \textbf{Large} & \textbf{Medium} & \textbf{large} \\
    \midrule
    SABRE  & 7.68 & 7.18 & 6.00 & 5.46 & 28.16 & 24.42 \\
    QMAP    & 6.85 & 6.31 & 5.15 & 4.96 & timeout & timeout \\
    Cirq        & 7.64 & 7.42  & 6.27 & 6.12 & 16.66 & 14.85 \\
    Pytket      & 9.99 & 9.03 & 6.47 & 5.89 & 37.21 & 30.93 \\
    Qlosure     & \textbf{5.72} & \textbf{5.45} & \textbf{4.41} & \textbf{4.08} & \textbf{14.94} & \textbf{13.45} \\
    \bottomrule
  \end{tabular}
\end{table}

\begin{table}[ht]
\vspace{-1ex}
  \TableFontSize
  \centering
  \caption{Summary of QUEKO benchmark results: average SWAP ratio (Existing mapping method SWAPs / Qlosure SWAPs).}
\vspace{-1ex}
  \label{tab:swaps_summary}
  \begin{tabular}{p{0.8cm} c c c c c c}
    \toprule
    \multirow{2}{*}{\textbf{Mapper}}
      & \multicolumn{2}{c}{\textbf{Sherbrooke}} 
      & \multicolumn{2}{c}{\textbf{Ankaa‑3}}
      & \multicolumn{2}{c}{\textbf{Sherbrooke-2X}}\\
    \cmidrule(lr){2-3} \cmidrule(lr){4-5} \cmidrule(lr){6-7} 
      & \textbf{Medium} & \textbf{Large} & \textbf{Medium}
      & \textbf{Large} & \textbf{Medium} & \textbf{large} \\
    \midrule
    SABRE  & 1.17 & 1.20 & 1.27 & 1.29 & 1.30 & 1.31 \\
    QMAP    & 1.81 & 1.85 & 2.14 & 2.18 & timeout & timeout \\
    Cirq        & 1.20 & 1.24 & 1.24 & 1.26 & 1.08 & 1.12 \\
    Pytket      & 1.32 & 1.29 & 1.23 & 1.24 & 1.42 & 1.37 \\
    \bottomrule
  \end{tabular}
  \shrink
  \vspace{-1ex}
\end{table}

Qlosure is designed 
to robustly improve 
mapping of circuits while keeping mapping times practical, 
as shown on the
queko-bss-54qbt dataset (Tab.\ref{tab:time_summary}). On 
Sherbrooke, Qlosure maps medium circuits in 6.07 s and large ones in
10.13 s, 
noticeable
faster than QMAP (10.36 s and 23.49 s), Cirq (5.85 s
and 13.14 s), and Pytket (14.54 s and 32.99 s). 
A similar 
edge appears on Ankaa-3.
On Sherbrooke-2X, Qlosure's mapping time grows from 7.36 s (on
medium) to 12.77 s for large circuits ($1.76\times$ growth), whereas SABRE grows
from 0.67 s to 1.77 s ($2.64\times$), QMAP from 11.48 s to 26.10 s
($2.27\times$), Cirq from 6.07 s to 13.48 s ($2.22\times$), and Pytket from
15.84 s to 37.95 s ($2.39\times$). 
Across all backends, our mapping time
grows by only 1.5–1.7$\times$ 
when moving from medium to large circuits, compared to
2.2–2.6$\times$ 
for the other mappers, 
highlighting 
that our method scales 
with circuit size. 
We also note that
LightSABRE, 
written
in Rust,
still 
yields 
faster 
times, but Qlosure strikes a superior balance of
quality and speed.

\begin{table}[h]
\vspace{-1ex}
  \TableFontSize
  \centering
  \caption{Average mapping times (in seconds) for QUEKO 54qbt dataset.
  }
  \vspace{-1ex}
  \label{tab:time_summary}
  \begin{tabular}{p{0.8cm} c c c c c c}
    \toprule
    \multirow{2}{*}{\textbf{Mapper}}
      & \multicolumn{2}{c}{\textbf{Sherbrooke}} 
      & \multicolumn{2}{c}{\textbf{Ankaa‑3}}
      & \multicolumn{2}{c}{\textbf{Sherbrooke-2X}}\\
    \cmidrule(lr){2-3} \cmidrule(lr){4-5} \cmidrule(lr){6-7} 
      & \textbf{Medium} & \textbf{Large} & \textbf{Medium}
      & \textbf{Large} & \textbf{Medium} & \textbf{large} \\
    \midrule
    SABRE  & 0.64 & 1.57 & 0.66 & 1.52 & 0.67 & 1.77 \\
    QMAP    & 10.36 & 23.49 & 8.45& 19.59 & 11.48 & 26.10 \\
    Cirq        & 5.85 & 13.14 & 4.56 & 9.89 & 6.07 & 13.48\\
    Pytket      & 14.54 & 32.99 & 9.49 & 20.90 & 15.84 & 37.95 \\
    Qlosure     & 6.07 & 10.13 & 4.07 & 6.09 & 7.36 & 12.77\\
    \bottomrule
  \end{tabular}
\vspace{-2ex}
\end{table}

\begin{figure}[tbh] 
    \centering
    \includegraphics[height=2.5cm,width=0.35\textwidth]{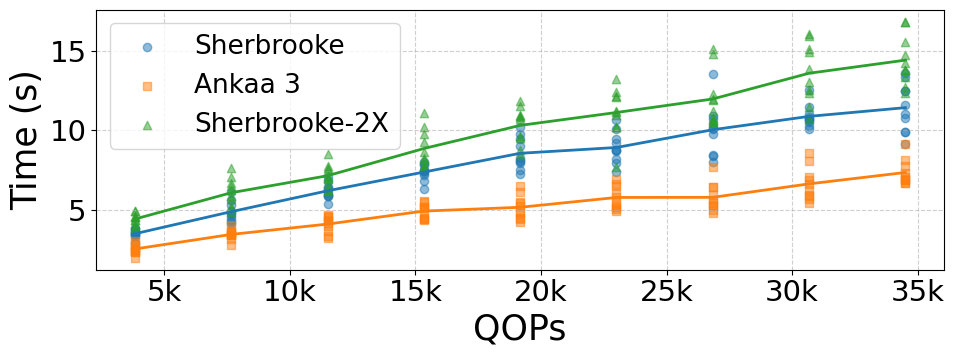} 
    \vspace{-2ex}
    \caption{Mapping time of Qlosure as a function of Quantum Operations (QOPs).
    }
    \label{fig:qosure_mapping_time_vs_qops}
\vspace{-2ex}
\end{figure}


\begin{figure*}[t]
  \centering
  \begin{subfigure}[b]{0.28\textwidth}
    \includegraphics[width=\linewidth]{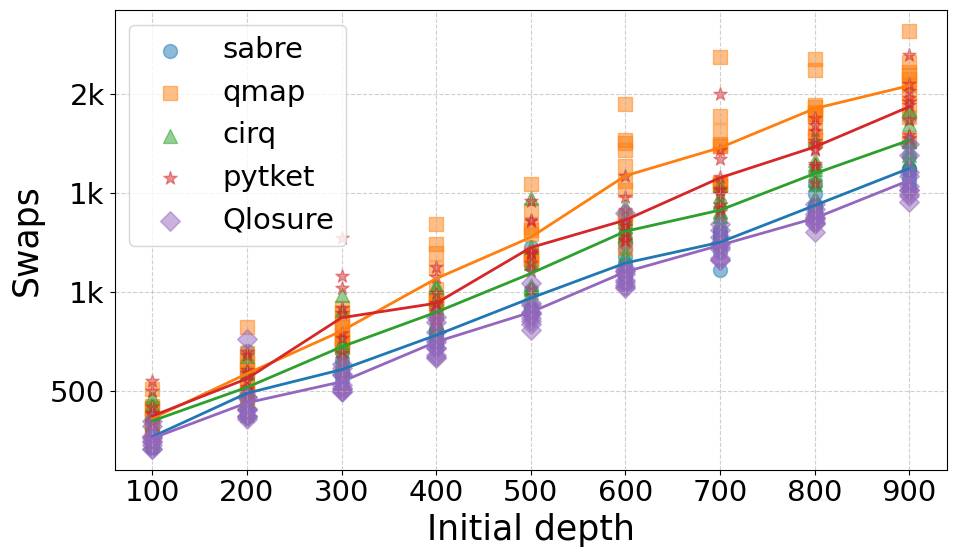}
    \vspace{-4ex}
    \caption{queko-bss-16qbt – Swaps}
  \end{subfigure}
  \begin{subfigure}[b]{0.28\textwidth}
    \includegraphics[width=\linewidth]{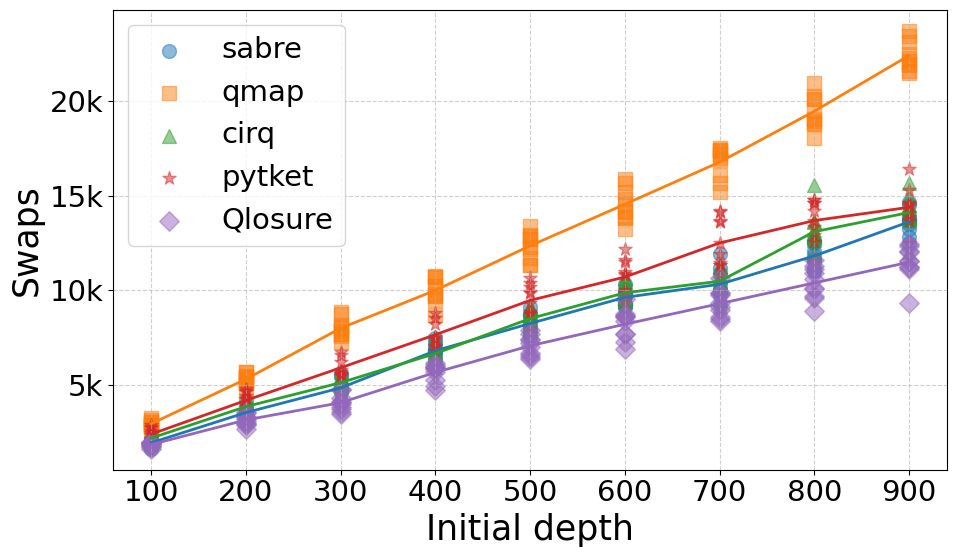}
       \vspace{-4ex}
    \caption{queko-bss-54qbt – Swaps}
  \end{subfigure}
  \begin{subfigure}[b]{0.28\textwidth}
    \includegraphics[width=\linewidth]{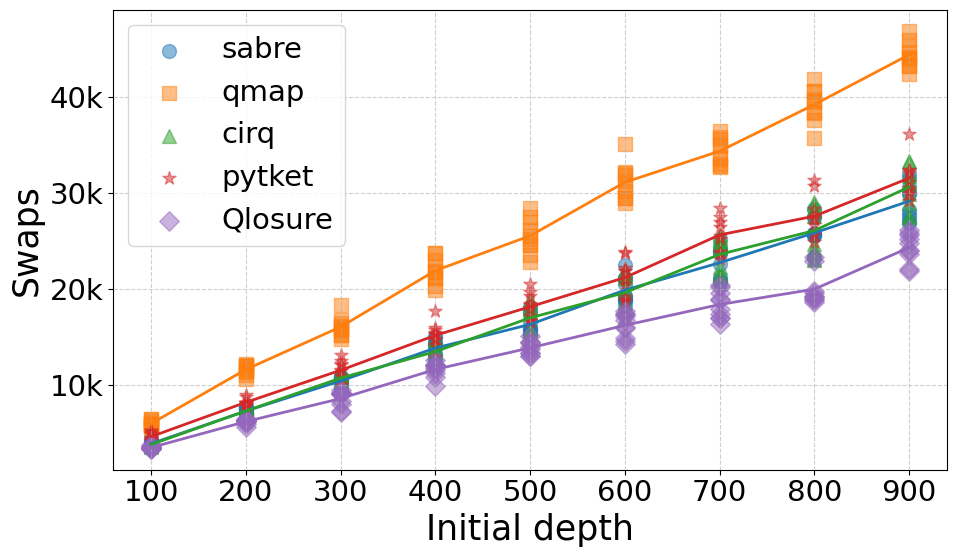}
       \vspace{-4ex}
    \caption{queko-bss-81qbt – Swaps}
  \end{subfigure}

  \begin{subfigure}[b]{0.28\textwidth}
    \includegraphics[width=\linewidth]{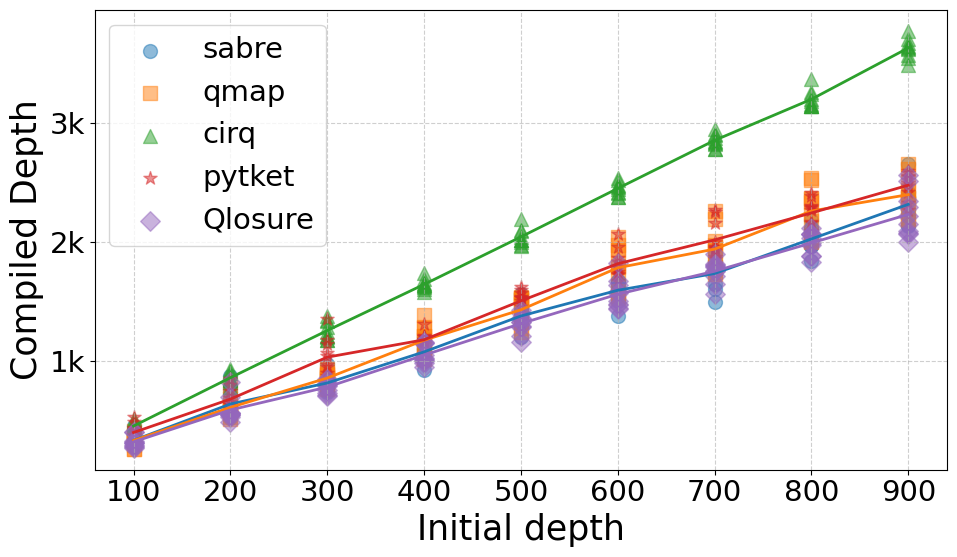}
       \vspace{-4ex}
    \caption{queko-bss-16qbt – Depth}
  \end{subfigure}
  \begin{subfigure}[b]{0.28\textwidth}
    \includegraphics[width=\linewidth]{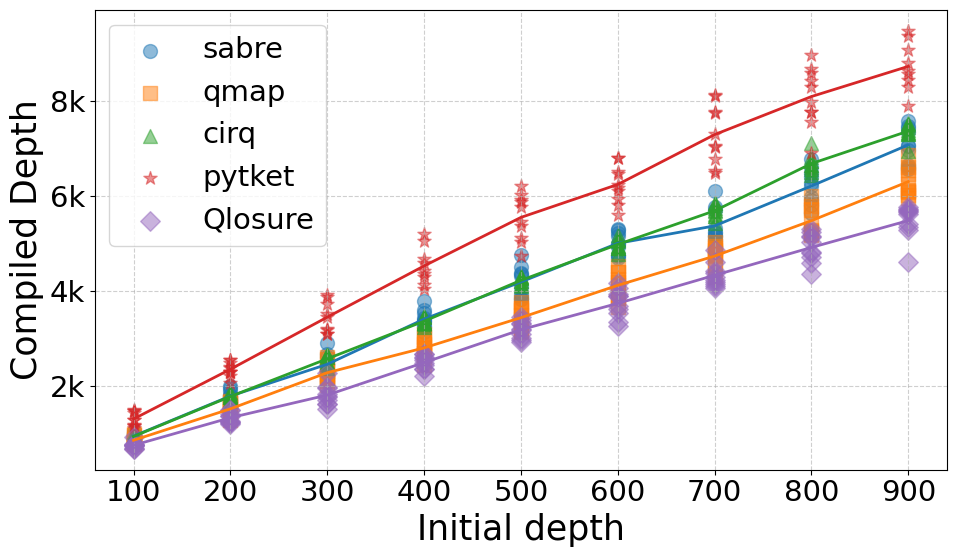}
       \vspace{-4ex}
    \caption{queko-bss-54qbt – Depth}
  \end{subfigure}
  \begin{subfigure}[b]{0.28\textwidth}
    \includegraphics[width=\linewidth]{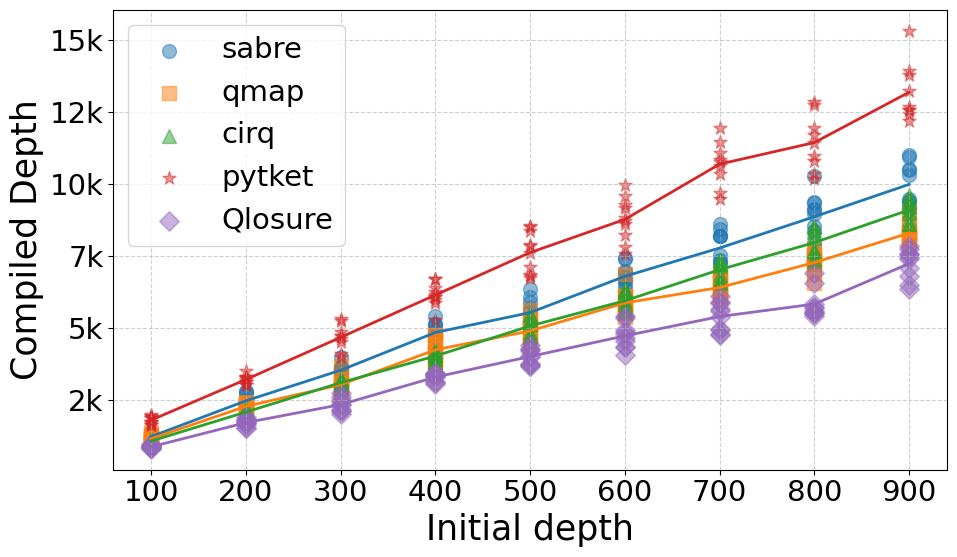}
       \vspace{-4ex}
    \caption{queko-bss-81qbt – Depth}
  \end{subfigure}
  \vspace{-1ex}
  \caption{Qubit mapping results on the {\bf Sherbrooke} backend for narrow, medium, and wide QUEKO benchmarks. Top: swap counts; bottom: circuit depths.
  %
  }
  \label{fig:sherbrooke}
  \shrink
\end{figure*}

Fig.~\ref{fig:qosure_mapping_time_vs_qops} further shows
Qlosure's scalability on the \texttt{queko-bss-54qbt} benchmark. 
For Sherbrooke, Ankaa-3, and Sherbrooke-2X backends, 
the mapping time 
grows
near-linearly with the number of Quantum Operations (QOPs). 
This predictable
scaling demonstrates that Qlosure’s dependence-driven heuristics offer
substantial circuit improvements while maintaining practical performance as
circuit complexity grows.

\subsection{Evaluation with QUEKO Circuits}
\label{sec:eval:queko}

We present here
detailed results on the QUEKO suite,
which permits us to perform comparisons
relative to known optimum circuit depth.
Metrics used are the number of inserted SWAPs and the circuit depth.
Experiments were conducted using 
the standard QUEKO queko-bss-16qbt (narrow) and queko-bss-54qbt (medium), 
and our custom-generated QUEKO instances (queko-bss-81qbt, wide) on the 
Sherbrooke and Ankaa-3 backends. 

\begin{figure*}[t]
  \centering
  \begin{subfigure}[b]{0.28\textwidth}
    \includegraphics[width=\linewidth]{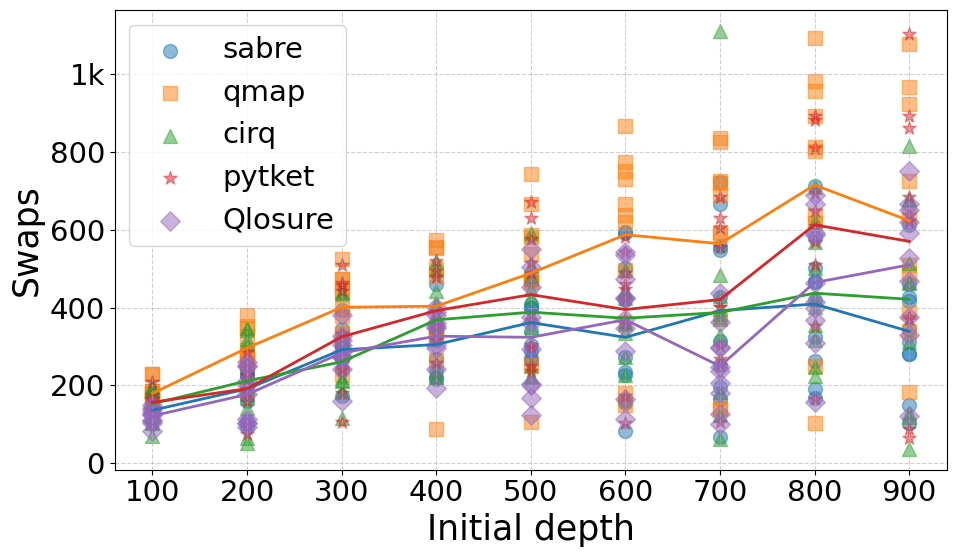}
    \shrink[2]
    \caption{queko-bss-16qbt – Swaps}
  \end{subfigure}
  \begin{subfigure}[b]{0.28\textwidth}
    \includegraphics[width=\linewidth]{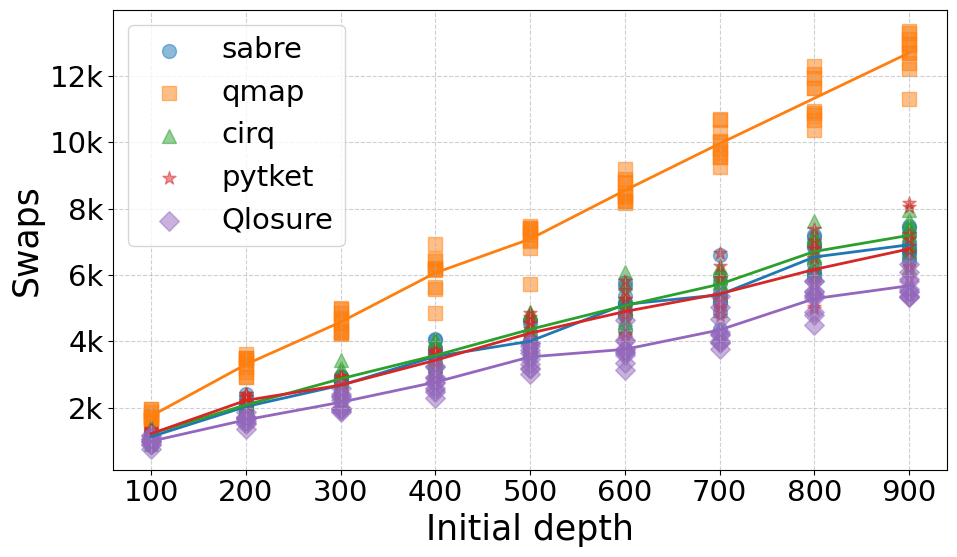}
    \shrink[2]
    \caption{queko-bss-54qbt – Swaps}
  \end{subfigure}
  \begin{subfigure}[b]{0.28\textwidth}
    \includegraphics[width=\linewidth]{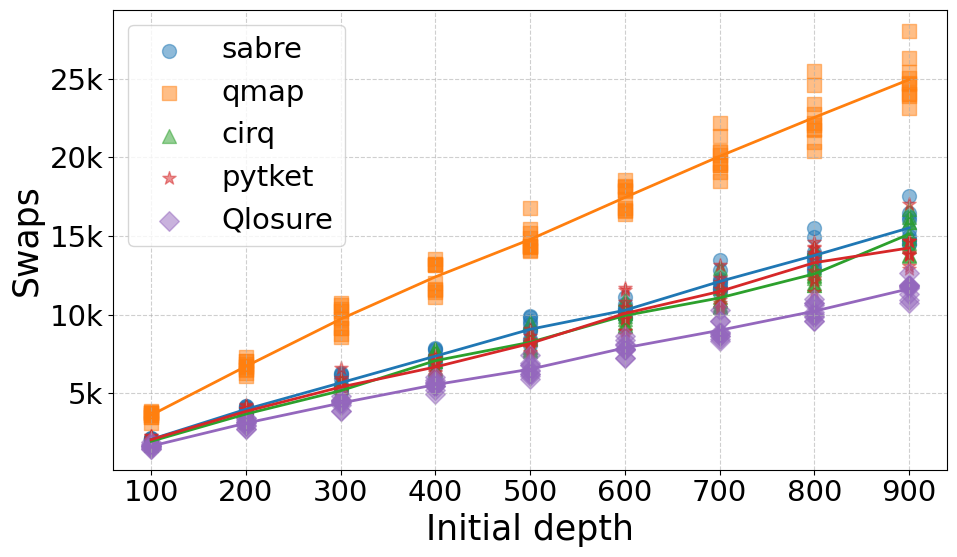}
    \shrink[2]
    \caption{queko-bss-81qbt – Swaps}
  \end{subfigure}

  \begin{subfigure}[b]{0.28\textwidth}
    \includegraphics[width=\linewidth]{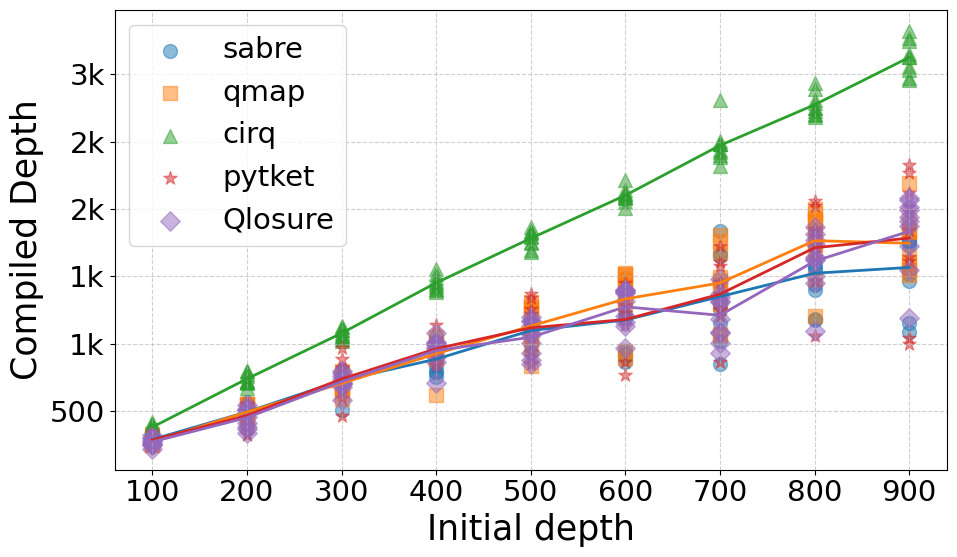}
    \shrink[2]
    \caption{queko-bss-16qbt – Depth}
  \end{subfigure}
  \begin{subfigure}[b]{0.28\textwidth}
    \includegraphics[width=\linewidth]{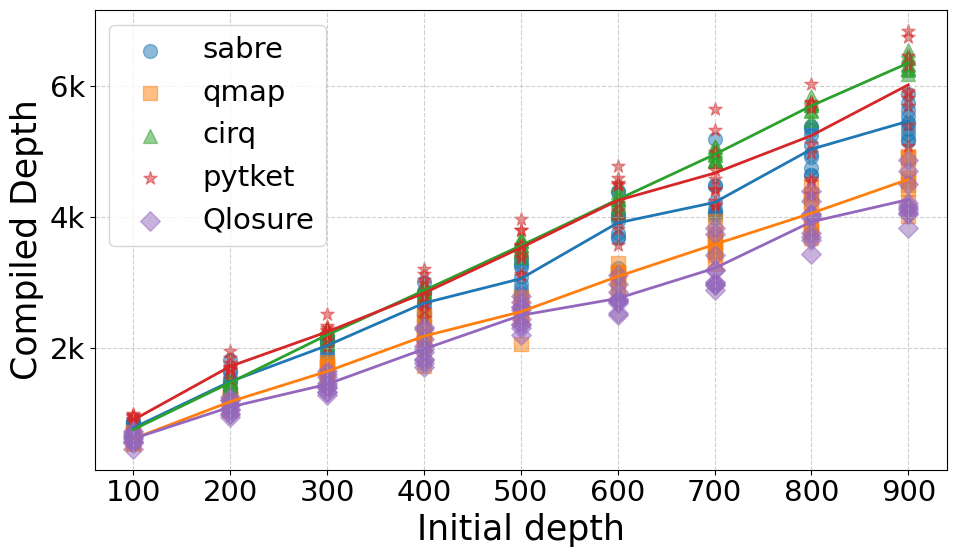}
    \shrink[2]
    \caption{queko-bss-54qbt – Depth}
  \end{subfigure}
  \begin{subfigure}[b]{0.28\textwidth}
    \includegraphics[width=\linewidth]{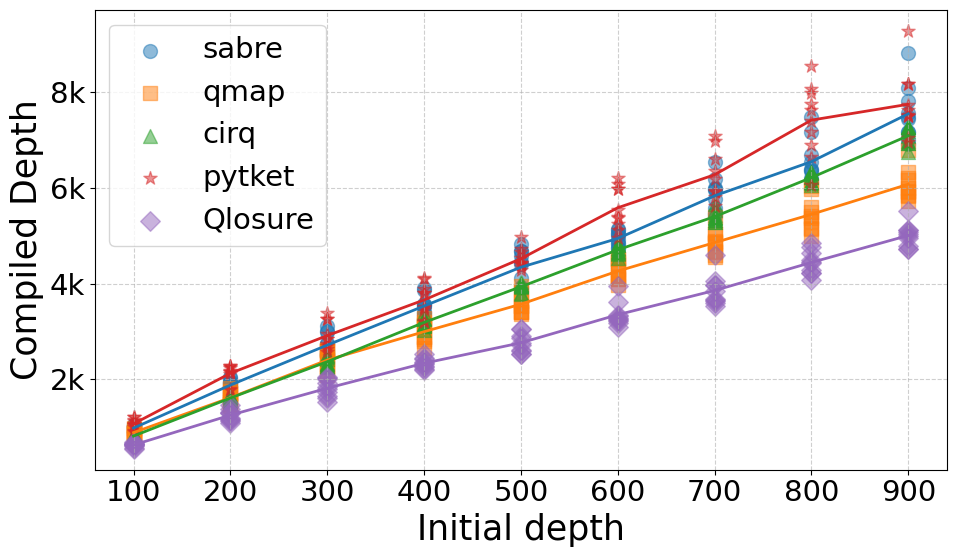}
    \shrink[2]
    \caption{queko-bss-81qbt – Depth}
  \end{subfigure}
  \vspace{-1ex}
  \caption{Qubit mapping results on the {\bf Ankaa-3} backend for narrow, medium, and wide QUEKO datasets. Top: swap counts; bottom: circuit depths.}
  \label{fig:ankaa}
\vspace{-2ex}
\end{figure*}

Fig.~\ref{fig:sherbrooke} shows the SWAP counts (top row) and final
circuit depths (bottom row) for each mapper 
on the Sherbrooke backend, across 
narrow
(queko-bss-16qbt), 
medium (queko-bss-54qbt), and 
wide
(queko-bss-81qbt) 
dataset
categories. 
Similarly, Fig.~\ref{fig:ankaa}
shows results on Ankaa-3. 
The
x-axis is the initial pre-mapping depth of 
circuits, 
allowing
for an analysis of how mappers scale with 
growing circuit size.

To ensure an equitable comparison, all mappers
start from an
identical, trivial (identity) initial mapping of logical to physical qubits.
Using the same initial mapping allows for an objective assessment of each
algorithm's intrinsic mapping capabilities.

{\bf Results on the Sherbrooke Backend}
(Fig.~\ref{fig:sherbrooke}):
On the smaller \texttt{queko-bss-16qbt} circuits, Qlosure 
outperformed
Sabre in 74.4\% of 
instances, while 
improving over
remaining
mappers in more than 94.4\%.
As circuit complexity increases with the \texttt{queko-bss-54qbt} and subsequently the \texttt{queko-bss-81qbt} datasets, 
Qlosure advantages became more pronounced. 
For instance, on the larger 81-qubit circuits, 
Qlosure achieves reductions in SWAP 
count ranging
from \textbf{17.95\%} (compared to Sabre) to \textbf{46.82\%} (compared to QMAP).
Similar reductions in final circuit depth were
also observed, with improvements ranging from \textbf{18.52\%} (over QMAP) to \textbf{49.22\%} (compared to Pytket).

{\bf Results on the Ankaa-3 Backend}
(Fig.~\ref{fig:ankaa}):
We organize our discussion by the dataset size
(in qubits).
On \texttt{queko-bss-16qbt} circuits (Fig.~\ref{fig:ankaa}a--\ref{fig:ankaa}d),
mappers
tend to yield comparable results,
because QUEKO-16 benchmarks
are generated targeting Rigetti Aspen-4, an architecture with a
topology similar to Rigetti Ankaa-3, which can reduce the apparent
differences between mapping strategies. 

\begin{table*}[t]
  \TableFontSize
  \centering
  \caption{Qubit mapping results on the {\bf Sherbrooke} backend for medium and large QasmBench datasets.
  Average improvement is computed as
  (VAL$_{baseline}$ - VAL$_{\ourtool{}}$)~/~VAL$_{baseline}$.
  Values in {\bf bold} denote empirical minimum value for circuit.
    \vspace{-1ex}
  }
  \label{tab:qasmbench-sherbrooke}
  \begin{tabular}{l r r *{5}{r r} r r}
    \toprule
    Circuit         & Qubits & QOPs 
      & \multicolumn{2}{c}{LightSABRE}
      & \multicolumn{2}{c}{MQT QMAP}
      & \multicolumn{2}{c}{Cirq}
      & \multicolumn{2}{c}{pytket}
      & \multicolumn{2}{c}{Qlosure}\\
    \cmidrule(lr){4-5} \cmidrule(lr){6-7}
    \cmidrule(lr){8-9} \cmidrule(lr){10-11}
    \cmidrule(lr){12-13} 
      &        & 
      & Swaps & Depth 
      & Swaps & Depth 
      & Swaps & Depth 
      & Swaps & Depth 
      & Swaps & Depth \\
    \midrule

qram\_n20        & 20 &  346   
  & 154   & \textbf{266}     
  & 173   & 316     
  & 175   & 305     
  & 176   & 292     
  & \textbf{145}   & \textbf{266}  \\

qugan\_n39       & 40 & 1036   
  & 197   &  416    
  & 193   &  664    
  & error    &   error    
  & 185   &  473    
  & \textbf{182}   & \textbf{347} \\

multiplier\_n45  & 45 & 5571   
  & 2318  & 3810    
  & 2230  & 4242    
  & 2060  & 4250    
  & 2355  & 3981    
  & \textbf{1923}  & \textbf{3679}  \\

qft\_n63         & 63 & 8689   
  & 2540  & 1998    
  & 2393  & 3021    
  & 3368  & \textbf{1524}    
  & 3299  & 2333    
  & \textbf{2313}  & 1831\\

adder\_n64       & 64 & 1156   
  & \textbf{593}   &  749    
  & 623& 1045 
  & 666   &  920    
  & 607   &  831    
  & 598   & \textbf{663}   \\

qugan\_n71       & 71 & 1932   
  & \textbf{525}   & 1366    
  & 530   & 1171    
  & 631   &  933    
  & 649   & 1292    
  & 558   & \textbf{904} \\

multiplier\_n75  & 75 &15767   
  & 5870  &10961    
  & 6390  &11177    
  & 6224  &12207    
  & 7416  &11936    
  & \textbf{5719}  & \textbf{10113}  \\
 \midrule
\multicolumn{3}{l}{Our average improvement (41 Circs)}
  & 7.40\% & 3.96\%
  & 11.89\% & 26.40\%
  & 13.31\% & 14.16\%
  & 14.28\% & 10.25\%
  &           &            \\
    \bottomrule
  \end{tabular}
\end{table*}

\begin{table*}[t]
  \TableFontSize
  \centering
  \caption{Qubit mapping results on the {\bf Ankaa-3} backend for medium and large QasmBench datasets. Table cell description identical to Tab.~\ref{tab:qasmbench-sherbrooke}.
  \vspace{-1ex}
  }
  \label{tab:qasmbench-ankaa}
  \begin{tabular}{l r r *{6}{r r} r r}
    \toprule
    Circuit & Qubits & QOPs
      & \multicolumn{2}{c}{LightSABRE}
      & \multicolumn{2}{c}{MQT QMAP}
      & \multicolumn{2}{c}{Cirq}
      & \multicolumn{2}{c}{pytket}
      & \multicolumn{2}{c}{Qlosure}\\
    \cmidrule(lr){4-5} \cmidrule(lr){6-7}
    \cmidrule(lr){8-9} \cmidrule(lr){10-11}
    \cmidrule(lr){12-13}
      &       & 
      & Swaps & Depth 
      & Swaps & Depth 
      & Swaps & Depth 
      & Swaps & Depth 
      & Swaps & Depth  \\
    \midrule

qram\_n20       & 20 & 346   
  &  80  &  231 
  &  78  &  256 
  & 102  &  258 
  &  81  &  222 
  & \textbf{70}  & \textbf{213}  \\

qugan\_n39      & 40 & 1036  
  & 164  &  578 
  & 157  &  619 
  & 185  &  428 
  & 160  &  456 
  & \textbf{149}  & \textbf{398} \\

multiplier\_n45 & 45 & 5571  
  &1416  & 3580 
  &1406  & 3705 
  &1674  & 3866 
  &1515  & 3531 
  & \textbf{1293}  & \textbf{3262} \\

qft\_n63        & 63 & 8689  
  &1705  & 1952 
  &2196  & 6513 
  &2136  & \textbf{1272} 
  &1837  & 1950 
  & \textbf{1549}  & 1713 \\

adder\_n64      & 64 & 1156  
  & 356  &  575 
  & 346  &  777 
  & 542  &  785 
  & 315  &  602 
  & \textbf{303}  & \textbf{566} \\

qugan\_n71      & 71 & 1932  
  & 465  &  916 
  & \textbf{310}  & 1051 
  & 433  &  814 
  & 427  &  923 
  & 367  & \textbf{803} \\

multiplier\_n75 & 75 &15767  
  &4325  & 9708 
  &4995  &10865 
  &4975  &11062 
  &4221  & 9936 
  & \textbf{4178}  & \textbf{9244} \\

    \midrule
\multicolumn{3}{l}{Our average improvement (41 Circs)}
  & 10.36\% & 5.59\%
  & 8.37\% & 27.95\%
  & 21.20\% & 15.46\%
  & 6.73\% & 5.96\%
  &           &            \\

    \bottomrule
  \end{tabular}
  \vspace{-3ex}
\end{table*}

In medium-sized circuits (\texttt{queko-bss-54qbt})
Qlosure's efficiency increases.
Specifically, on SWAP insertions, Qlosure shows
superior results in
\textbf{94\%} of instances 
over baseline methods.
Regarding final circuit depths, 
Qlosure edges
QMAP in \textbf{71.1\%} of
cases and 
all other 
mappers 
in
\textbf{100\%} of instances.

On \texttt{queko-bss-81qbt} circuits,
our results further improve,
in average
yielding
the lowest SWAP count and circuit depth. 
On these larger inputs, Qlosure reduced SWAP gates
by margins from \textbf{19.87\%} (over Cirq) to \textbf{54.66\%} (relative to QMAP). 
Moreover, improvements
in final circuit depths 
were also significant,
ranging
between \textbf{20.79\%} (compared to QMAP) and \textbf{38.90\%} (compared to Pytket).


\vspace{-1ex}
\subsection{Evaluation with QasmBench Circuits}
\label{sec:eval:qasmbench}
Next, we present our results on QasmBench, which
consists of practical, near-term 
circuits. 
We also note that QasmBench circuits are typically more sparse
than those of QUEKO. 
 As before, we organize our discussion by back-ends.
All QasmBench circuits with 20--81 qubits are used.
Tables show excerpts of results and a summary row of all 41 circuits.

\textbf{Sherbrooke}  
(Tab.\ref{tab:qasmbench-sherbrooke}): 
We
improve in both SWAP count and
circuit depth over other methods. 
For the 75-qubit \texttt{multiplier\_n75},
Qlosure delivers 2.5\% to 22\% fewer SWAPs and 7\% to 17\% smaller
depth, compared to the other mappers (2.5\% fewer SWAPs compared to LightSABRE, and 22\% compared to pytket).
On the 39-qubit \texttt{qugan\_n39}, we
obtain
reductions of 2\%–7\% and  16\%–46\%, for SWAP and depth, respectively. 
Averaged
over all 41 circuits, Qlosure yields 7.4\% fewer SWAPs and 3.96\% shorter depth
versus LightSABRE, with even larger gains on other mappers.\\

\textbf{Ankaa-3}  
(Tab.\ref{tab:qasmbench-ankaa}): 
Overall results on Ankaa-3 
are better than on Sherbrooke,
and improve over all baselines.
For
the 75-qubit \texttt{multiplier\_n75}, Qlosure reduces SWAPs by 4\%–19\% and depth by
3\%–34\%.
On the 39-qubit \texttt{qugan\_n39}, 
SWAPs reduction range from 5\%–19\% and depth gains
of 7\%–36\%. 
Averaged over all 41
circuits, \ourtool{}
achieves a 10.4\% reduction in SWAPs and a 5.6\%
reduction in depth versus LightSABRE, with average gains of 8.4\% and 28.0\% over
QMAP, 
21.2\% and 15.5\% versus Cirq, and 6.7\% and 6.0\% against pytket.

\vspace{-1ex}
\subsection{Ablation Study: Impact of Cost Function Components}
\label{sec:eval:ablation}
\vspace{-1ex}
We use the \texttt{queko-bss-81qbt} dataset and Sherbrooke back-end
for this study.
Our goal is to dissect the impact of cost function factors
and trace the bulk of improvements to 
weights introduced for transitive gate
dependencies (see Eq.~\ref{eq:H}).

\begin{figure}[htb]
\begin{subfigure}{0.48\linewidth}
\includegraphics[height=3.05cm,width=\linewidth]{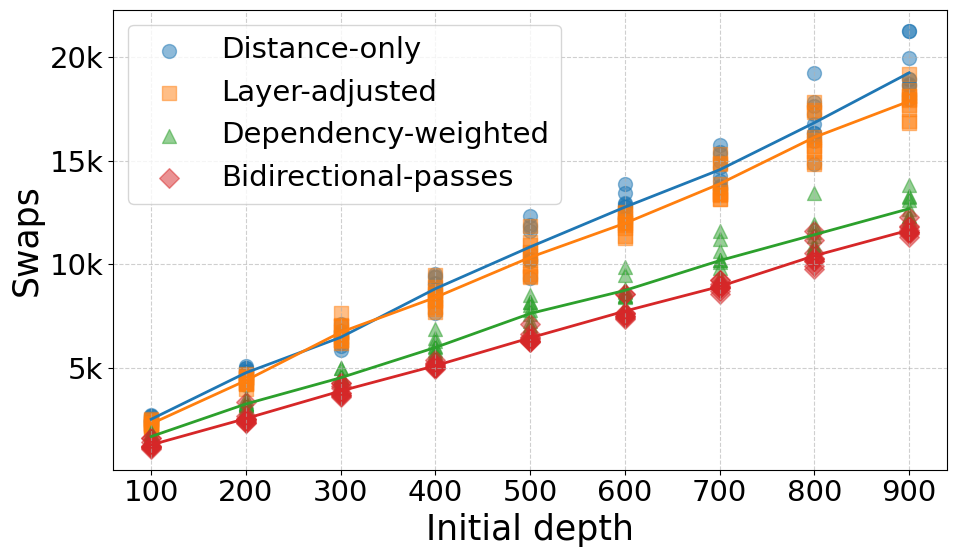} 
\vspace{-3ex}
\caption{Metric: SWAP Count.}
\label{fig:ablation_swaps_sherbrooke_81qbt}
\end{subfigure}
\begin{subfigure}{0.48\linewidth}
\centering
\includegraphics[height=3.05cm,width=\linewidth]{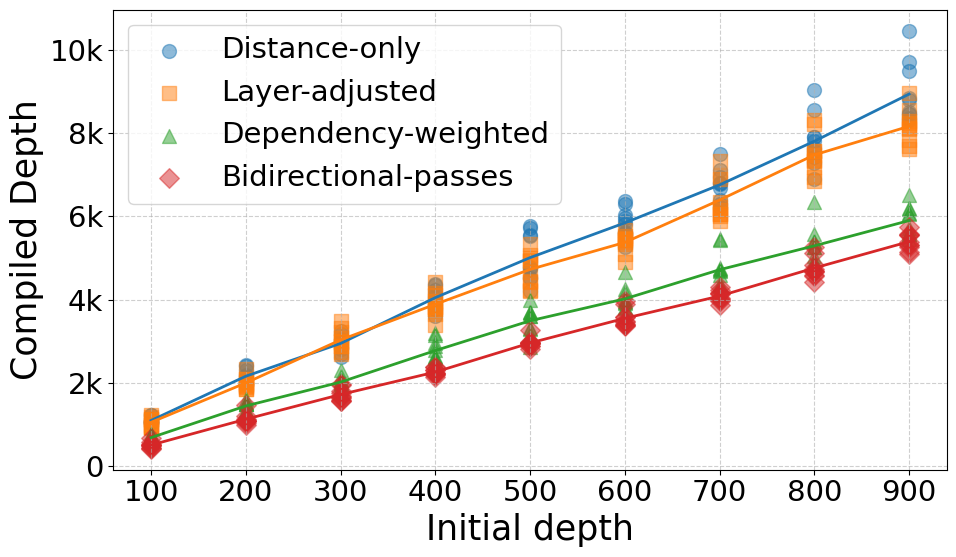} 
\vspace{-3ex}
\caption{Metric: Circuit depth.}
\label{fig:ablation_depth_sherbrooke_81qbt}
\end{subfigure}
 \vspace{-1ex}
 \caption{Ablation Study on Sherbrooke (`queko-bss-81qbt`).}
\vspace{-4ex}
\end{figure}

Fig.~\ref{fig:ablation_swaps_sherbrooke_81qbt}--\ref{fig:ablation_depth_sherbrooke_81qbt} plot the SWAP counts and resulting
depths as functions of the initial circuit depth.  We compare four variants,
with their average performance relative to the distance‐only baseline:
a) \textbf{Distance‐only (baseline)}, 
uses only the Manhattan distance in swap choices, setting 
our reference level.
b) \textbf{Layer‐adjusted},
adds layer‐based weights slightly favoring earlier gates, leading to 5.6\% fewer SWAPs and 5.9\% smaller depth.
c) \textbf{Dependency-weighted},
adds transitive dependence weights, yielding 46.8\% fewer SWAPs and 48.7\% smaller depth compared to the baseline.
d) \textbf{Bidirectional-passes},  
enhances our results by
using a
non-trivial initial qubit mapping 
obtained from a forward-backward
pass over the input circuit, similar to \cite{sabre}. 
Leveraging the non-trivial mapping 
results in 72.2\% fewer SWAPs and 76.8\%
smaller depth.
We remark that results of all previous sections use the trivial mapping as initial placement.




\vspace{-5ex}
\noindent

\section{Related Work}
\label{sec:related}






Various methods 
for circuit mapping 
on nearest-neighbor devices 
have been proposed. We discuss relevant work next.


QMAP \cite{mqt-qmap-jku-wille-burgholzer.ispd.2023} 
implements 
exact and
heuristic algorithms
targeting
specific hardware configurations
\cite{qmap-optimal-subarch-qmap.tqc.2023} 
via 
subarchitecture search 
in quantum devices, while
also aiming to prune
large search spaces 
using exact mapping strategies 
\cite{search-space-qc-qmap-jku.aspdac.2022}.

As
scalability 
is a key concern
in NISQ devices, several heuristics have been proposed.
Park et al. \cite{fast-scalable-qubit-mapping.dac.2022}
introduced a scalable qubit mapping method designed to reduce
compilation overhead on such systems. 
On modular quantum architectures, Baker
et al. \cite{time-slicing-circuit-partitioning.cf.2020} explored time-sliced
partitioning of quantum circuits
to map 
sub-circuits and manage inter-module communication. 
Looking towards future
hardware, Escofet et al. \cite{revisiting-mapping-qc.tqc.2025} 
revisited 
the mapping problem in the context of emerging multi-core quantum processors,
observing new challenges and opportunities. Another approach to
scalability, ``route-forcing''
\cite{route-forcing-scalable-quantum-circuit-mapping.qce.2024}, 
aimed to
map circuits to large-scale quantum 
devices
by 
judiciously
constraining routing choices.

Machine learning (ML) has also been applied to quantum circuit mapping. Fan et al. \cite{quantum-circ-placement-machine-learning-ml-qc.dac.2022} leveraged ML to optimize quantum circuit placement, a crucial sub-problem of mapping. Similarly, Paler et al. \cite{quantum-circuit-layout-opt-with-ml.tqc.2023} applied ML for the broader optimization of quantum circuit layouts to improve performance on physical devices. Beyond direct mapping, Xu et al. \cite{synthesis-qc-optimizers.pldi.2023} explored the automatic synthesis of quantum circuit optimizers, which could potentially generate novel mapping strategies.

Prior work have
also targeted device and program features.
Molavi et al.
\cite{dependence-aware-compilation-qec-annealing.oopsla.2025} focused on
dependence-aware compilation for {\em surface code architectures}, 
noting
the role of data dependencies in optimizing for error correction. 
Niu et al.
\cite{qubit-mapping-hardware-aware.tqe.2020} presented a hardware-aware
heuristic for qubit mapping tailored to the constraints of NISQ devices. 
For
exact solutions on smaller systems, Molavi et al.
\cite{qubit-mapping-routing-with-maxsat.micro.2022} 
proposed a MaxSAT 
qubit mapping and routing
that delivered
scalable
results
on QPUs with typically fewer than 20 qubits. Steinberg et al.
\cite{topo-graph-dependencies-qubit-mapping-heuristic.tqe.2022} 
investigated the
influence of topological graph dependencies within a heuristic qubit assignment
algorithm and analyzed its scaling properties. 
Finally, Banerjee et al.
\cite{locality-aware-qubit-mapping.ipdpsw.2022} proposed a locality-aware qubit
routing strategy,
customized
to grid-based quantum architectures,
aiming to minimize 
SWAP gate insertions.

\vspace{-1ex}
\section{Conclusion}
\label{sec:conclusion}
\vspace{-1ex}
The main scientific contribution of our work
is the methodology to harness
affine dependence relations, and in particular,
transitive dependences, to make critical decisions when 
introducing SWAP operations. The advantage of transitive
dependences is that they capture the effect
of candidate SWAPs on future gates, 
allowing 
us to make decisions that affect
fewer gates along the circuit critical path.
The algorithm leverages a look-ahead window of
near-term gates, further organized
into layers, and makes critical decisions considering
all transitive dependences originating in the look-ahead window.
We validate our approach on three quantum devices while
comparing against four state-of-the-art qubit mappers,
yielding circuit depth improvements of up to 2.5$\times$
and 40\% fewer SWAPs.
Possible future research directions are
jointly exploiting dependence information and
device topology together with
customized qubit-state and error-aware mapping heuristics.



\vspace{-1ex}
\section{Acknowledgment}
This research has been partly supported by the Center for Quantum and Topological Systems at New York University Abu Dhabi. Experiments were carried out on the High-Performance Computing resources at New York University Abu Dhabi.


\newcommand{\ranswer}[1]{\textcolor{red}{#1}}
\newcommand{\fixme}[1]{\textcolor{red}{\underline{\bf #1}}}

\bibliographystyle{IEEEtran}
\bibliography{bib/quantum,bib/martin,bib/polyhedral}

\appendix

\subsection{More Detailed Background}

\subsubsection{Integer Sets and Relations}
\label{appendix:sets}

In this section, we provide an overview of two main concepts used in the polyhedral model: \emph{integer sets} and \emph{maps}. These two concepts are used in the polyhedral model to represent code, code transformations, and to reason and analyze code.

\emph{Integer sets} represent iteration domains while \emph{maps} are used to represent accesses to qubits and to transform iteration domains and access relations.

An integer set is a set of integer tuples described using affine constraints.  An example of a set of integer tuples is {$\{(1,1); (2,1); (3,1); (1,2); (2,2); (3,2)\}$}.
Instead of listing all the tuples as we do in the previous set, we can describe the set using affine constraints over loop iterators and symbolic constants as follows: {$\{S(i,j): 1 \leq i \leq 3 \wedge 1 \leq j \leq 2\}$} where $i$ and $j$ are the dimensions of the tuples in the set.

Formally, an integer set is defined as:
$S = \{\, \mathbf{i} \in \mathbb{Z}^n \mid \varphi(\mathbf{i})\,\}$.
Here, \( \mathbf{i} = (i_1, i_2, \ldots, i_n) \) is an integer vector, and \( \varphi(\mathbf{i}) \) is a Presburger formula—a logical expression constructed from linear inequalities, logical connectives (such as \( \land, \lor, \lnot \)), and quantifiers (like \( \forall, \exists \)), without involving multiplication of variables.

A map is a relation between two integer sets.  For example
{$\{S1(i,j) \rightarrow S2(i+2,j+2) : 1 \leq i \leq 3 \wedge 1 \leq j \leq 2\}$}
is a map between tuples in the set S1 and tuples in the set S2 (e.g. the tuple {$S1(i,j)$} maps to the tuple {$S2(i+2,j+2)$}).

All sets and maps in this work are implemented using the Integer Set Library (ISL)~\cite{isl}. We also use the ISL library notation for sets and maps throughout the paper.

\subsubsection{Transitive Closure}
\label{appendix:transitive}

\begin{figure}[h]
  \centering
  \includegraphics[width=0.5\linewidth]{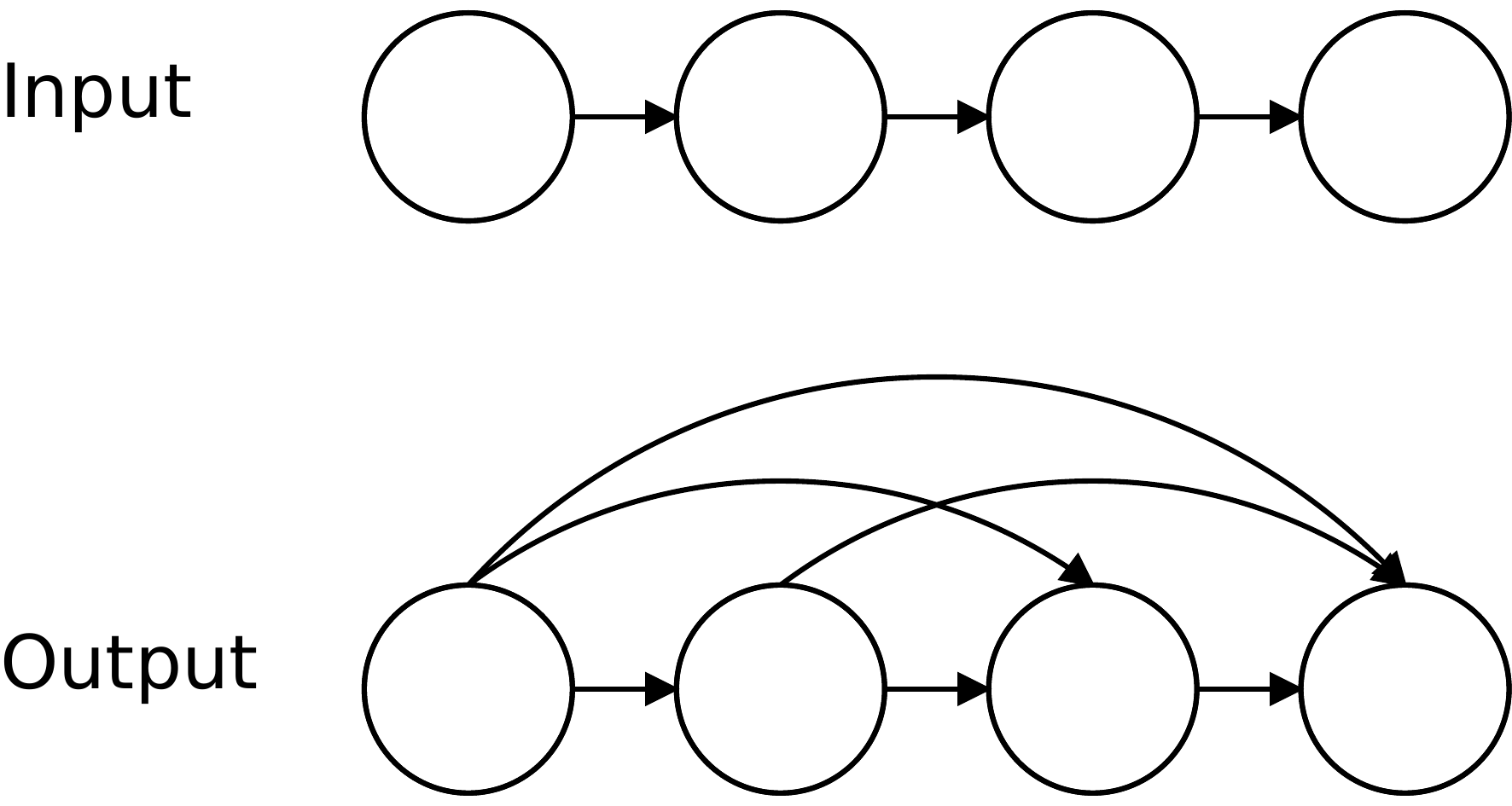}
  \caption{Transitive closure constructs the output dependence graph below from the input point-to-point dependence graph (top).}
  \label{fig:transitive}
\end{figure}

Figure~\ref{fig:transitive} shows an example of applying the transitive closure on an input graph. The output graph is shown in the figure. Using the transitive closure, we can answer reachability questions. For example, can we get from node a to node c? A binary relation tells us only that node a is connected to node b, and that node b is connected to node c. After the transitive closure is constructed, in an O(1) operation, one can determine that node c is reachable from node a.

The transitive closure is formally defined as:

{
\FormulaSize
\begin{equation*}
R^{+} = \bigcup_{l} R^l, \text{ with } R^l 
\begin{cases}
R & \text{ if } l = 1 \\
R \circ R^{l-1} & \text{ if } l > 1 \\
\end{cases}
\end{equation*}
}

\subsubsection{Qubits}
\label{appendix:qubit}
A qubit's state, often denoted as \(\lvert \psi \rangle\), can be \(\lvert 0 \rangle\), \(\lvert 1 \rangle\), or a linear combination (superposition) of these two basis states: \(\lvert \psi \rangle = \alpha \lvert 0 \rangle + \beta \lvert 1 \rangle\). The coefficients \(\alpha, \beta \in \mathbb{C}\) are complex numbers satisfying the normalization condition \(|\alpha|^2 + |\beta|^2 = 1\), where \(|\alpha|^2\) is the probability of measuring the qubit in state \(\lvert 0 \rangle\), and \(|\beta|^2\) is the probability of measuring \(\lvert 1 \rangle\).
In addition, qubits can also be \emph{entangled}. For example, the general state of a two-qubit system is  $\lvert \Psi \rangle = \alpha_{00} \lvert 00 \rangle + \alpha_{01} \lvert 01 \rangle + \alpha_{10} \lvert 10 \rangle + \alpha_{11} \lvert 11 \rangle$, this is called \emph{entanglement} where the two-qubit state \(\lvert \Psi \rangle\) cannot be written as a simple product of individual qubit states, implying correlated measurement outcomes.

\subsection{More Details on the Experimental Setup}

\subsubsection{Quantum Hardware Backends}
\label{appendix:backends}

\begin{figure}[ht!]
  \centering
  \begin{subfigure}[b]{0.24\textwidth}
    \centering
    \includegraphics[width=\linewidth]{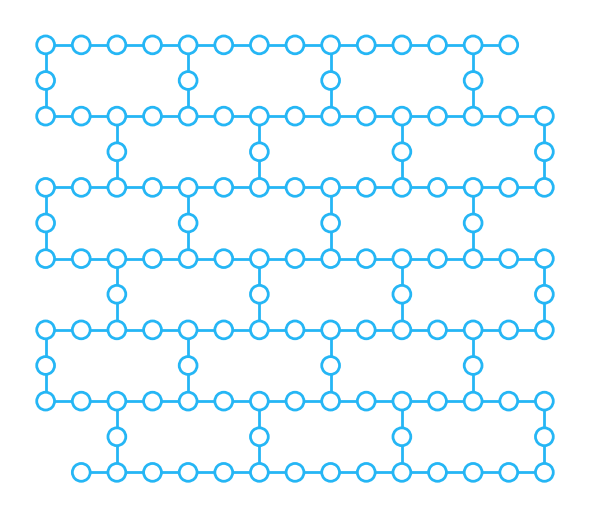}
    \caption{IBM Sherbrooke.}
    \label{fig:sherbrooke_topo}
  \end{subfigure}
  \hfill
  \begin{subfigure}[b]{0.24\textwidth}
    \centering
    \includegraphics[width=\linewidth,height=4cm]{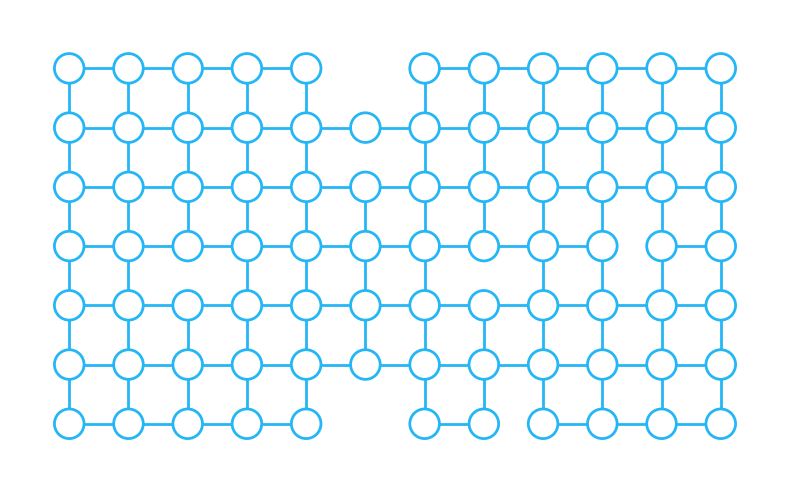}
    \caption{Rigetti Ankaa 3.}
    \label{fig:ankaa_topo}
  \end{subfigure}
  \begin{subfigure}[b]{0.48\textwidth}
    \centering
    \includegraphics[width=\linewidth]{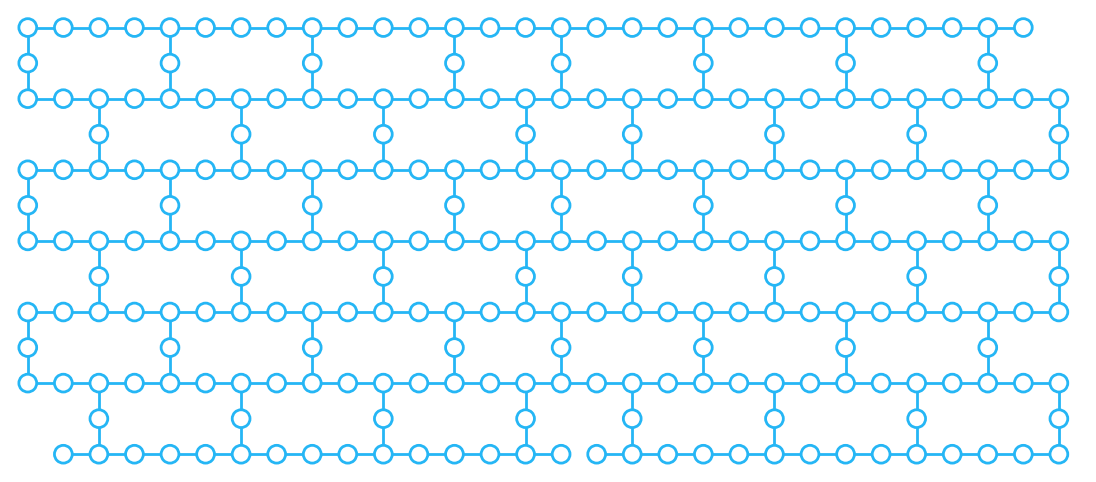}
    \caption{IBM Sherbrooke 2X.}
    \label{fig:sherbrooke2x_topo}
  \end{subfigure}
  \caption{Coupling graphs.}
  \label{fig:topologies}
\end{figure}

Figure \ref{fig:topologies} show coupling graphs for the IBM Sherbrooke, Rigetti Ankaa-3 and  
Sherbrooke-2X quantum hardware backends used in our experiments.

\end{document}